\documentclass{aa}
\usepackage{graphicx}
\usepackage{bm,natbib,url,wasysym}
\graphicspath{{./fig/}{./png/}}
\newcommand{\EQ}{\begin{equation}}
\newcommand{\EN}{\end{equation}}
\newcommand{\EQA}{\begin{eqnarray}}
\newcommand{\ENA}{\end{eqnarray}}

\newcommand{\Eq}[1]{Equation~(\ref{#1})}
\newcommand{\Eqs}[2]{Equations~(\ref{#1}) and~(\ref{#2})}

\newcommand{\Sec}[1]{Section~\ref{#1}}

\newcommand{\Fig}[1]{Figure~\ref{#1}}

\newcommand{\Figs}[2]{Figures~\ref{#1} and \ref{#2}}

\newcommand{\Tab}[1]{Table~\ref{#1}}

\newcommand{\bra}[1]{\langle #1\rangle}

\newcommand{\meanrho}{\overline{\rho}}

{}
{}
{}

{}
{}
{}
{}
{}
{}
{}
{}
\newcommand{\meanAA}{\overline{\mbox{\boldmath $A$}}{}}{}
\newcommand{\meanBB}{\overline{\mbox{\boldmath $B$}}{}}{}
\newcommand{\meanEE}{\overline{\mbox{\boldmath $E$}}{}}{}
{}
\newcommand{\meanFFf}{\overline{\mbox{\boldmath $F$}}_{\rm f}{}}{}
{}
{}
{}
{}
{}

{}
{}
{}
\newcommand{\meanA}{\overline{A}}
\newcommand{\meanB}{\overline{B}}

\newcommand{\meanF}{\overline{F}}

\newcommand{\meanhf}{\overline{h}_{\rm f}}

{}

{}
{}

\newcommand{\Rsun}{R}
\newcommand{\eee}{\hat{\mbox{\boldmath $e$}} {}}

\newcommand{\gggg}{\mbox{\boldmath $g$} {}}

\newcommand{\rr}{\mbox{\boldmath $r$} {}}

\newcommand{\kk}{\bm{k}}

\newcommand{\xx}{\bm{x}}

\newcommand{\uu}{\mbox{\boldmath $u$} {}}
\newcommand{\UU}{\mbox{\boldmath $U$} {}}

\newcommand{\BB}{\mbox{\boldmath $B$} {}}
\newcommand{\EE}{\mbox{\boldmath $E$} {}}

\newcommand{\JJ}{\mbox{\boldmath $J$} {}}

\newcommand{\AAA}{\mbox{\boldmath $A$} {}}
\newcommand{\aaaa}{\mbox{\boldmath $a$} {}}
\newcommand{\ee}{\mbox{\boldmath $e$} {}}

\newcommand{\ff}{\mbox{\boldmath $f$} {}}

\newcommand{\FF}{\mbox{\boldmath $F$} {}}

\newcommand{\nab}{\mbox{\boldmath $\nabla$} {}}

\newcommand{\RRRR}{\mbox{\boldmath ${\sf R}$} {}}
\newcommand{\SSSS}{\mbox{\boldmath ${\sf S}$} {}}

\newcommand{\ii}{{\rm i}}
\newcommand{\erf}{{\rm erf}}

\newcommand{\DD}{{\rm D} {}}

\newcommand{\dd}{{\rm d} {}}

\newcommand{\const}{{\rm const}  {}}

\def\Pm{P_{\rm m}}
\def\Rm{R_{\rm m}}

\def\cs{c_{\rm s}}

\def\vA{v_{\rm A}}

\def\kmean{k_{\rm m}}

\def\kf{k_{\rm f}}
\def\Af{A_{\rm f}}

\def\Brms{B_{\rm rms}}

\def\urms{u_{\rm rms}}

\def\kappah{\kappa_{\rm h}}
\def\kappat{\kappa_{\rm t}}

\def\etat{\eta_{\rm t}}
\def\etatz{\eta_{\rm t0}}

\def\Beq{B_{\rm eq}}

\def\half{{\textstyle{1\over2}}}

\def\onethird{{\textstyle{1\over3}}}

\newcommand{\G}{\,{\rm G}}

\newcommand{\s}{\,{\rm s}}

\newcommand{\cm}{\,{\rm cm}}

\newcommand{\m}{\,{\rm m}}

\newcommand{\kms}{\,{\rm km/s}}

\newcommand{\kg}{\,{\rm kg}}

\newcommand{\Mm}{\,{\rm Mm}}
\newcommand{\Mx}{\,{\rm Mx}}

\newcommand{\yjgr}[3]{ #1, {J.\ Geophys.\ Res.,} {#2}, #3}

\newcommand{\yapj}[3]{ #1, {ApJ,} {#2}, #3}

\newcommand{\yapjl}[3]{ #1, {ApJ,} {#2}, #3}

\newcommand{\yan}[3]{ #1, {Astron.\ Nachr.,} {#2}, #3}

\newcommand{\yana}[3]{ #1, {A\&A,} {#2}, #3}

\newcommand{\ygafd}[3]{ #1, {Geophys.\ Astrophys.\ Fluid Dyn.,} {#2}, #3}
\newcommand{\ygrl}[3]{ #1, {Geophys.\ Res.\ Lett.,} {#2}, #3}

\newcommand{\ypf}[3]{ #1, {Phys.\ Fluids,} {#2}, #3}

\newcommand{\ypp}[3]{ #1, {Phys.\ Plasmas,} {#2}, #3}

\newcommand{\yprl}[3]{ #1, {Phys.\ Rev.\ Lett.,} {#2}, #3}

\newcommand{\ymn}[3]{ #1, {MNRAS,} {#2}, #3}

\newcommand{\ysph}[3]{ #1, {Sol. Phys.,} {#2}, #3}

\newcommand{\ypre}[3]{ #1, {Phys.\ Rev.\ E,} {#2}, #3}

\newcommand{\yjour}[4]{ #1, {#2}, {#3}, #4}

\newcommand{\ybook}[3]{ #1, {#2} (#3)}

\begin{document}

\titlerunning{Dynamo-driven plasmoid ejections above a spherical surface}
\authorrunning{J. Warnecke et al.}

\title{Dynamo-driven plasmoid ejections above a spherical surface}
\author{J\"orn Warnecke\inst{1,2} \and Axel Brandenburg\inst{1,2} \and
  Dhrubaditya Mitra\inst{1}}
\institute{Nordita, AlbaNova University Center, Roslagstullsbacken 23,
SE-10691 Stockholm, Sweden
\and Department of Astronomy, AlbaNova University Center,
Stockholm University, SE-10691 Stockholm, Sweden}

\abstract{}{%
We extend earlier models of turbulent dynamos with an upper,
nearly force-free exterior to spherical geometry, and study how
flux emerges from lower layers to the upper ones without being
driven by magnetic buoyancy.
We also study how this affects the possibility of plasmoid ejection.
}{%
A spherical wedge is used that includes northern and southern hemispheres
up to mid-latitudes and a certain range in longitude of the Sun.
In radius, we cover both the region that corresponds to the convection
zone in the Sun and the immediate exterior up to twice the radius of the
Sun.
Turbulence is driven with a helical forcing function in the interior,
where the sign changes at the equator between the two hemispheres.
}{%
An oscillatory large-scale dynamo with equatorward migration
is found to operate in the turbulence zone.
Plasmoid ejections occur in regular intervals, similar to what is
seen in earlier Cartesian models.
These plasmoid ejections are tentatively associated with 
coronal mass ejections (CMEs).
The magnetic helicity is found to change sign outside the turbulence
zone, which is in agreement with recent findings for the solar wind.
}{}
\keywords{Magnetohydrodynamics (MHD) -- turbulence --
Sun: dynamo -- Sun: coronal mass ejection (CMEs) -- 
stars: magnetic fields
}

\maketitle

\section{Introduction}

Observations show that the Sun sheds mass through twisted magnetic
flux configurations \citep{Demoulin}.
Remarkable examples of such helical ejections 
can be seen in the movies produced by the SOHO and SDO missions\footnote{
\url{http://sohowww.nascom.nasa.gov/bestofsoho/Movies/10th/transcut_sm.mpg}
and
\url{http://www.youtube.com/watch?v=CvRj6Uykois&feature=player_embedded}}.
Such events may be important for the solar dynamo \citep{BB03}.
They are generally referred to as coronal mass ejections (CMEs). 
Conventionally, CMEs are modeled by adopting a given 
distribution of magnetic flux at the solar surface and letting it 
evolve by shearing and twisting the magnetic field at its footpoints 
at the surface \citep{ADK99,TK03}.
This approach is also used to model coronal heating
\citep{GN05,Bingert}.
The success of this method depends crucially on the ability to
synthesize the velocity and magnetic field patterns at the surface.
Of course, ultimately such velocity and magnetic field patterns must
come from a realistic simulation of the Sun's convection zone, where
the field is generated by dynamo action.
In other words, we need a unified treatment of the convection zone
and the CMEs. 
The difficulty here is the large range of time scales, from the 11-year
dynamo cycle to the time scales of hours and even minutes on which
CMEs develop.
Such a large range of time scales is related to the
strong density stratification in the Sun, 
as can be seen from the following argument. 
In the bulk of the convection zone, the dynamo is controlled by rather
slow motions with turnover times of days and months.
The typical velocity depends on the convective flux via
$F_{\rm conv}\approx\meanrho\urms^3$, where $\meanrho$ is the mean
density and $\urms$ is the rms velocity of the turbulent convection.
The dynamo cycle time can even be several hundred times the turnover time.
In the corona, on the other hand, the typical time scale depends on
the Alfv\'en time, $L/\vA$, where $L$ is the typical scale of magnetic
structures and $\vA=B/\sqrt{\mu_0\meanrho}$ is the Alfv\'en speed
for a given magnetic field strength $B$.
Here, $\mu_0$ is the vacuum permeability.

In a recent paper, \cite{WB10} attempted
a new approach of a unified treatment by combining a dynamo-generated field in the
convection zone with a nearly force-free coronal part, albeit
in a local Cartesian geometry.
In this paper, we go a step further by performing direct numerical simulations (DNS) 
in spherical geometry.
We also include density stratification due to gravity, but with a
density contrast between the dynamo interior and the outer parts of the
simulation domain that is much less (about 20) than in the Sun
(around 14 orders of magnitude).
This low density contrast is achieved by using an isothermal configuration
with constant sound speed $\cs$.
Hence, the average density depends only on the gravitational potential
and is given by $\ln\rho(r)\approx GM/r\cs^2$, where $G$ is Newton's
gravitational constant, $M$ is the central mass,
and $r$ is the distance from the center.
As convection is not possible in such an isothermal setup, 
we drive turbulence by an imposed helical forcing that
vanishes outside the convection zone.
This also helps achieving a strong large-scale magnetic field.
The helicity of the forcing is negative (positive) in the 
northern (southern) hemisphere and smoothly changes sign 
across the equator.
Such a forcing gives rise to an $\alpha^2$
dynamo with periodic oscillations and equatorward migration of
magnetic activity \citep{Mitra10}. 
We ignore differential rotation, so there
is no systematic shearing in latitude.
The only twisting comes then from the same motions that also sustain the
dynamo-generated magnetic field.
Note that in our model the mechanism of transport of the magnetic field to the
surface is not magnetic buoyancy. 
Instead, we expect that, twisted magnetic fields will expel
themselves to the outer regions by the Lorentz force.

Our aim in this paper is not to provide a model as close to reality as
possible, but to show that it is possible to capture the phenomenon of CMEs
(or, more generally, plasmoid ejections)
within a minimalistic model that treats the convection
zone and the outer parts of the Sun in a self-consistent manner. 
That is, the magnetic field in the convection zone is dynamically
generated by dynamo action and the motions are not prescribed by
hand, but they emerge as a solution of the momentum equation and include
magnetic backreaction from the Lorentz force.

Given that gravity decreases with radius, there is in principle the
possibility of a radial wind with a critical point at $r_*=GM/2\cs^2$
\citep{Cho98}.
However, as we use stress-free boundary conditions with no mass flux in
the radial direction, no such wind can be generated in our simulations. 
Nevertheless, we observe radially outward propagation of helical magnetic
field structures without mass flux.
Furthermore, our results for the flux of magnetic helicity compare well  
with recent measurements of the same in the solar wind \citep{BSBG11}.
Our approach might therefore provide new insights not only for CMEs and
dynamo theory, but also for solar wind  turbulence.

\section{The model}
\label{TheModel}

We use spherical polar coordinates, $(r,\theta,\phi)$.
As in earlier work of \cite{Mitra09,Mitra10}, our simulation domain
is a spherical wedge.
The results of \cite{Mitra09} for such a wedge are consistent with those
of \cite{LHT10} for a full spherical shell, both of which ignored the
effects of an equator, which was included in the work of \cite{Mitra10}.
Our model is a bi-layer in the radial direction. 
The inner layer is forced with random helical forcing functions
which have different signs of helicity in the two hemispheres. 
This models the helical aspects of convection in the Sun. 
We shall often call the inner layer ``turbulence zone''. 
The radius separating the two layers corresponds to the solar
radius, $r=R$. 
This length scale is used as our unit of length.  
The inner layer models some aspects of the convection zone
($0.7\Rsun\leq r\leq\Rsun$) without however having any real convection,
and the outer layer ($\Rsun\leq r\leq2\Rsun$) models the solar corona.
We consider the range $\pi/3\leq\theta\leq2\pi/3$, corresponding to
$\pm30^\circ$ latitude, and $0<\phi<0.3$, corresponding to a
longitudinal extent of $17^\circ$.
Here, $\theta$ is the polar angle and $\phi$ the azimuth. 
At the solar surface at $R=700\Mm$, this would correspond to an area of
about $730\times210\Mm^2$, which could encompass several active regions
in the Sun.
In our model the momentum equation is
\begin{equation}
{\DD\UU\over\DD t}= -\nab h + \gggg
+\JJ\times\BB/\rho+\FF_{\rm for}+\FF_{\rm visc},
\label{DUDtext}
\end {equation}
where $\FF_{\rm visc}=\rho^{-1}\nab\cdot(2\rho\nu\SSSS)$ is the viscous
force, $\nu$ is the kinematic viscosity,
${\mathsf S}_{ij}=\half(U_{i;j}+U_{j;i})-\onethird\delta_{ij}\nab\cdot\UU$
is the traceless rate-of-strain tensor, semicolons denote covariant
differentiation, $h=\cs^2\ln\rho$ is the specific pseudo-enthalpy,
$\cs=\const$ is the isothermal sound speed, and
$\gggg=-GM\rr/r^3$ is the gravitational acceleration.
We choose $GM/\Rsun\cs^2=3$, so $r_*=1.5\Rsun$ lies within our domain.
This value is rather close to the surface and would lead to significant
mass loss if there was a wind, but this is suppressed by using
impenetrative outer boundaries.

The forcing function $\FF_{\rm for}$ is given as the product of two
parts,
\begin{equation}
\FF_{\rm for}(r,\theta,\phi,t)=\Theta_w(r-R)\,\ff(r,\theta,\phi,t;-\cos\theta),
\label{ForcingFunction}
\end{equation}
where $\Theta_w(r)=\half\left[1-\erf{\left(r/w\right)}\right]$
is a profile function connecting the two layers and $w$ is the width of the
transition at the border between the two layers ($r=\Rsun$).
In other words, we choose the external force to be zero in the 
outer layer, $r>\Rsun$. 
The function $\ff$ consists of random plane helical
transverse waves with relative helicity
$\sigma=(\ff \cdot \nab \times \ff)/k_{\rm f} \ff^2$ and wavenumbers
that lie in a band around an average forcing wavenumber of $\kf R\approx63$.
This value should be compared with the normalized wavenumber $k_1 R$,
corresponding to the thickness of the shell $\Delta R$, which yields
$k_1 R=2\pi R/\Delta R\approx21$, so the effective scale separation
ratio, $\kf/k_1$, is about 3.

In \Eq{ForcingFunction} the last argument of
$\ff(r,\theta,\phi,t;\sigma)$ denotes a parametric
dependence on the helicity which is here chosen
to be $\sigma=-\cos\theta$ such that the kinetic helicity of the turbulence
is negative in the northern hemisphere and positive in the southern.
Specifically, $\bm f$ is given by \citep{Hau04}
\begin{equation}
\ff(\xx,t)=\Af N{\rm Re}\{\ff_{\kk(t)}\exp[\ii\kk(t)\cdot\xx+\ii\phi(t)]\},
\label{Af}
\end{equation}
where $\Af$ is a nondimensional forcing amplitude,
and $\xx$ is the position vector.
The wavevector $\kk(t)$ and the random phase $-\pi<\phi(t)\le\pi$ change
at every time step, so $\ff(\xx,t)$ is $\delta$-correlated in time.
The normalization factor $N$ is chosen on dimensional grounds to be
$N=c_{\rm s}(|\kk|c_{\rm s}/\delta t)^{1/2}$.
At each timestep we select randomly one of many possible wavevectors
in a certain range around a given forcing wavenumber.
The average wavenumber is referred to as $k_{\rm f}$.
We ignore curvature effects in the expression for the forcing function
and thus force the system with what would correspond to transverse helical
waves in a Cartesian coordinate system, i.e.,
\begin{equation}
\ff_{\kk}=\RRRR\cdot\ff_{\kk}^{\rm(nohel)}\quad\mbox{with}\quad
{\sf R}_{ij}={\delta_{ij}-\ii\sigma\epsilon_{ijk}\hat{k}_k
\over\sqrt{1+\sigma^2}},
\end{equation}
where $-1\leq\sigma\leq1$ is the helicity parameter of the forcing function,
\EQ
\ff_{\kk}^{\rm(nohel)}=
\left(\kk\times\eee\right)/\sqrt{\kk^2-(\kk\cdot\eee)^2},
\label{nohel_forcing}
\EN
is a non-helical forcing function, and $\eee$ is an arbitrary unit vector
not aligned with $\kk$; note that $|\ff_{\kk}|^2=1$.

The pseudo-enthalpy term in \Eq{DUDtext} emerges from the fact that
for an isothermal
equation of state the pressure is given by $p=\cs^2\rho$, so the pressure
gradient force is given by $\rho^{-1}\nab p=\cs^2\nab\ln\rho=\nab h$.
The continuity equation is then written in terms of $h$ as
\begin{equation}
{\DD h\over\DD t}=-c_s^2\nab\cdot\UU.
\label{dhdt}
\end{equation}
\Eqs{DUDtext}{dhdt} are solved together with the uncurled induction
equation for the vector potential $\AAA$
in the resistive gauge \citep{CHBM11},
\begin{equation}
{\partial\AAA\over\partial t}=\UU\times\BB+\eta\nabla^2\AAA,
\end{equation}
where $\eta$ is the (constant) magnetic diffusivity,
so the magnetic field is given by $\BB=\nab\times\AAA$ and thus obeys
$\nab\cdot\BB=0$ at all times.
The gauge can in principle become important when calculating the magnetic
helicity density $\AAA\cdot\BB$, although the
part resulting from the small-scale fields is expected to be independent
of the gauge \citep{SB06,HB10}, while that of the large-scale fields
is not.

Our wedge is periodic in the azimuthal direction.
For the velocity, we use stress-free boundary conditions 
on all other boundaries.
For the magnetic field we employ vertical field conditions on
$r=2\Rsun$ and perfect conductor conditions on both $r=0.7\Rsun$ and
the two $\theta$ boundaries.
Time is expressed in units of $\tau = \left(\urms\kf\right)^{-1}$, which is
the eddy turnover time in the turbulence zone,
and $\urms$ is the rms velocity in $r<\Rsun$.
Density is given in units of the mean density in the turbulence zone,
$\rho_0=\meanrho$.
The magnetic field is expressed in units of the mean equipartition
value, $\Beq$, defined via $\Beq^2=\mu_0\overline{\rho\uu^2}$.
We use a magnetic diffusivity that is constant in space and time and its
value is given in terms of the magnetic Reynolds number, defined as
\begin{equation}
\Rm=\urms/\eta\kf,
\label{magRey}
\end{equation}
where $\kf$ is the characteristic scale of the external force, defined above.
This also turns out to be the energy containing scale of the fluid.  
In the following analysis, we use $\phi$ averages, defined as
$\overline{F}(r,\theta,t)$=$\int {F(r,\theta,\phi,t)\;\dd\phi/2\pi}$.
Occasionally we also use time averages denoted by $\bra{.}_t$.
We perform DNS with the 
{\sc Pencil Code}\footnote{\texttt{http://pencil-code.googlecode.com}},
which is
a modular high-order code (sixth order in space and third-order
in time, by default) for solving a large range of partial differential
equations, including the ones relevant in the present context.

\section{Results}
\subsection{Dynamo in the turbulence zone}
\label{Dynamo_TZ}

We start our DNS with seed magnetic field everywhere in the domain. 
Owing to the helical forcing in the turbulent layer, a large-scale
magnetic field is produced by dynamo action.
The dynamo is cyclic with equatorward migration of magnetic fields. 
This dynamo was studied by DNS in \cite{Mitra10} and has been
interpreted as an $\alpha^2$ dynamo.
The possibility of oscillating $\alpha^2$ dynamos was known since the
early papers of \cite{BS87} and \cite{RB87}, who showed that
a necessary condition for oscillations is that the 
$\alpha$ effect must change sign in the domain.   

\begin{figure}[t!]
\begin{center}
\includegraphics[width=\columnwidth]{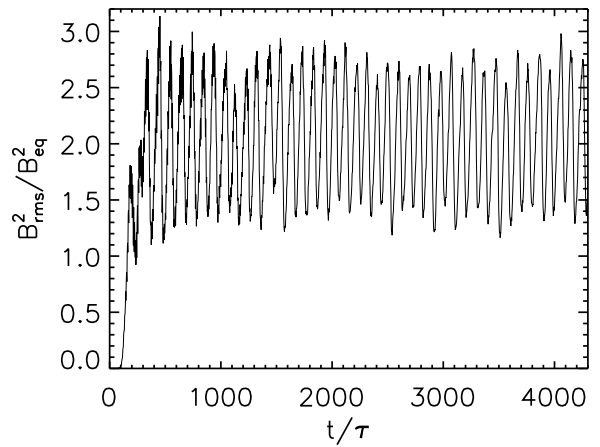}
\end{center}\caption[]{
Initial exponential growth and subsequent saturation behavior of
the magnetic field in the interior for forced turbulence with dynamo
action for Run~A.
The magnetic field strength is oscillating with twice the dynamo
frequency $2\omega_{\rm cyc}$.
}
\label{bbdym}
\end{figure}

\begin{figure}[t!]
\includegraphics[width=\columnwidth]{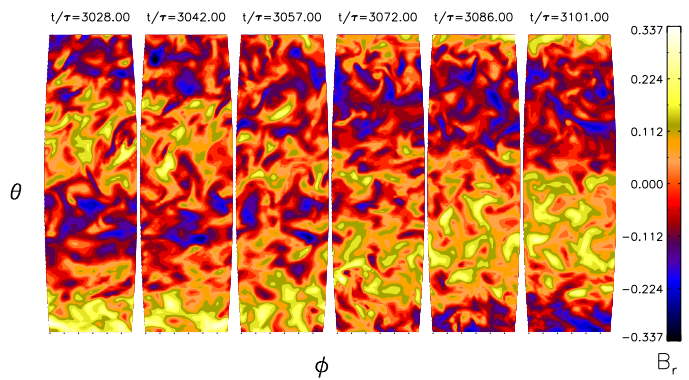}
\caption[]{
Equatorward migration, as seen in visualizations of $B_{r}$ for Run~D at $r=R$
over a horizontal extent $\Delta\theta=58^\circ$ and $\Delta\phi=17^\circ$.
}
\label{photoB}
\end{figure}

\begin{figure}[t!]
\begin{center}
\includegraphics[width=\columnwidth]{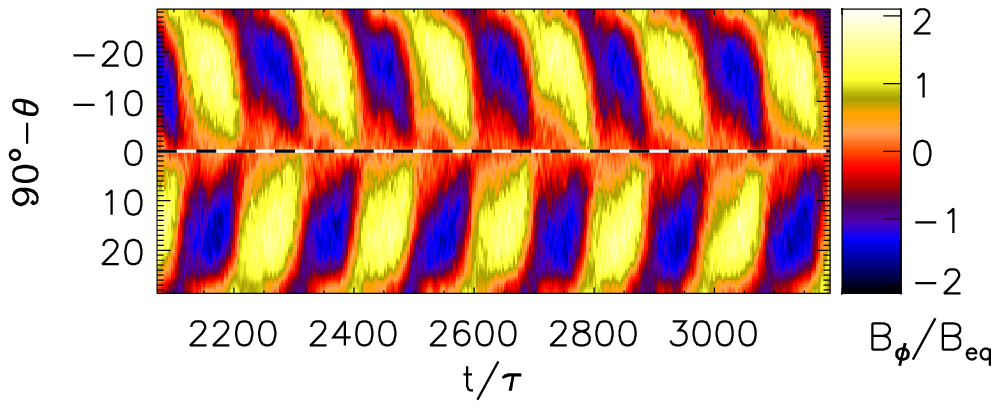}
\includegraphics[width=\columnwidth]{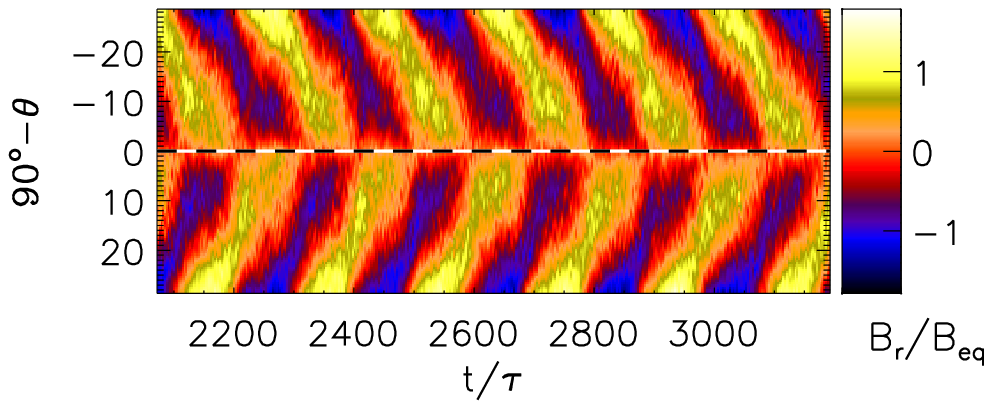}
\end{center}\caption[]{
Periodic variation of $\bra{\meanB_{\phi}}_r$ and $\bra{\meanB_{r}}_r$ 
in the turbulence zone.
Dark blue stands for negative and light yellow for positive values.
The dotted horizontal lines show the location of the equator at
$\theta=\pi/2$.
The magnetic field is normalized by the equipartition value.
Taken from Run~A.
}
\label{but2}
\end{figure}

\begin{table}[b!]\caption{
Summary of runs discussed in this paper.
}\vspace{12pt}\centerline{\begin{tabular}{lccccccccccc}
Run & 
$\Af$ & $\Rm$  & $\Pm$ & $\Brms^2/\Beq^2$ &$\tau \omega_{\rm cyc}$ &
$\Delta t/\tau$  &  $V_{\rm ej}/\urms$ \\
\hline
A  & 0.20 & 1.5 & 1 & 1.2--2.7& 0.032& 100& 0.482\\
B & 0.20 & 5 & 1 & 1.5--3.5& 0.029& 110& 0.409\\	  
C & 0.20 & 9 & 1 & 2.1--5.5& 0.022& 130& 0.455\\     
D & 0.20 & 18  &1 & 2.0--5.0& 0.019& 140& 0.409\\
D1 & 0.10 & 11  &1 & 2.0--5.0&0.018& 140& 0.455\\  
D2 & 0.15 & 15  &1 & 2.0--5.0&0.016 & 130& 0.482\\ 
D3 & 0.25 & 20  &1 & 1.0--3.0&0.023& 130&0.293\\ 
E & 0.20 & 22  &1 & 1.5--4.5& 0.017& 220 & 0.205\\ 
F & 0.20 & 28  &1 & 1.2--4.2& 0.015& 280& 0.273\\ 
G & 0.20 & 44  &1 & 1.7--3.5& 0.015& 285& 0.409\\ 
\label{Summary}\end{tabular}
}\tablefoot{
$\Af$ is the forcing amplitude defined in \Eq{Af},
$\Rm$ is the magnetic Reynolds number defined in \Eq{magRey}
and $\Pm=\nu/\eta$ is the magnetic Prandtl number.
$\omega_{\rm cyc}=2\pi/T_{\rm cyc}$ stands for the frequency of the
oscillating dynamo, where $T_{\rm cyc}$ is the cycle period.
$\Delta t/\tau$ is the typical interval of
plasmoid ejections, whose typical speed is $V_{\rm ej}$.
For the runs D1 to D3, the forcing amplitude $\Af$ is changed, while
$\eta$ and $\nu$ have the same value as for run D.
The rms velocities in the turbulence zone change accordingly
and affect therefore the Reynolds number and the turnover times $\tau$.
}\end{table}

The dynamo first grows exponentially and then saturates
after around 300 turnover times, see \Fig{bbdym}.
After saturation the dynamo produces a large-scale magnetic field with 
opposite polarities in the northern and southern hemispheres.
In \Fig{photoB} we plot the radial magnetic field at the surface of
the dynamo region at $r=R$.
This layer would correspond to the solar photosphere if we had a more
realistic solar model, which would include higher stratification
as well as cooling and radiation effects.
The six wedges represent different times and show clearly an
equatorward migration of the radial magnetic field with polarity
reversal every half cycle.
The other components of the magnetic field (not plotted) also shows
the same behavior. 
Comparing the first ($t/\tau=3028$) and the last ($t/\tau=3101$)
panel,
the polarity has changed sign in a time interval $\Delta
t/\tau\approx100$. 
The oscillatory and migratory properties of the dynamo is also seen
in the butterfly diagram of \Fig{but2} for 
$\bra{\meanB_{\phi}}_{r}$ and $\bra{\meanB_{r}}_{r}$.
In \Fig{bbdym} one can also verify that
the oscillation period is around 200
turnover times, corresponding to a non-dimensional dynamo cycle frequency of
$\tau\omega_{\rm cyc}=0.032$ and the field strength in the
turbulent layer varies between 1.2 and 1.6 of the equipartition field
strength.
This value of the cycle frequency is roughly consistent with an estimate
of \cite{Mitra10b} that $\omega_{\rm cyc}=0.5\etat\kmean^2$, where
$\kmean$ is the relevant wavenumber of the mean field.
Using $\etat\approx\etatz\equiv\urms/3\kf$ \citep{Sur08}, we find
$\tau\omega_{\rm cyc}\approx0.2(\kmean/\kf)^2\approx0.02$,
where we have assumed $\kmean\approx2\pi/0.3R\approx20k_1$
and $\kf\approx60k_1$.
The estimate of \cite{Mitra10b} applies to perfectly conducting outer
boundary conditions, which might explain the remaining discrepancy.

A summary of all runs is given in \Tab{Summary}, where the
amplitudes of the magnetic field show a weak non-monotonous dependence
on the magnetic Reynolds number $\Rm$.
For larger values of $\Rm$, the magnetic field strength decreases slightly with
increasing $\Rm$, but it is weaker than in some earlier $\alpha^2$ dynamos
with open boundaries \citep{BD01}.
This could be due to two reasons.
Firstly, our simulations are far from the asymptotic limit of
large magnetic Reynolds numbers,
in which the results of \cite{BD01} are applicable. 
The maximum value $\Rm$ is in our simulations
approximately 15 times the critical $\Rm$.
The second reason could be that we have expulsion of magnetic helicity
from our domain which was not present in \citep{BD01}.  
We find the peak of the $\Rm$ dependency at $\Rm=9$, corresponding to
Run~C. 
The dynamo cycle frequency shows a weak decrease (by a factor of 1.5) as
the magnetic Reynolds number increases (by a factor of 20).

\subsection{Phase relation between radial and azimuthal fields}

\begin{figure}[t!]
\begin{center}
\includegraphics[width=\columnwidth]{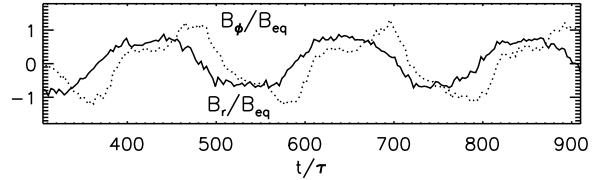}
\end{center}\caption[]{
Time evolution of the radial magnetic field $B_{r}$ (solid line) and
the azimuthal magnetic field $B_{\phi}$ (dotted line)
in the dynamo region averaged over the radius $r$ and azimuth
at $\theta$=$\pm7^{\circ}$.
To improve the statistics, we calculate the components of the
magnetic field as the antisymmetric part in latitude, i.e.,
$B_i=\left(B^{\rm N}_i-B^{\rm S}_i\right)/2$ for $i=r,\phi$.
}
\label{bphase}
\end{figure}

Although our dynamo model does not include important features of the Sun
such as differential rotation, some comparison may still be appropriate.
For the Sun, one measures the mean radial field by averaging
the line-of-sight magnetic field from synoptic magnetograms.
The azimuthal field is not directly observed, but its sign can normally
be read off by looking at the magnetic field orientation of sunspot pairs.
Existing data suggest that radial and azimuthal fields are approximately in
out-of-phase \citep{Yos76}.
This is reasonably well reproduced by $\alpha\Omega$ dynamos models, where
the radial field lags behind the azimuthal one by $0.75\pi$ \citep{Stix76}.
However, in the present work, radial and azimuthal fields are
approximately in phase with a phase difference of $0.3\pi$ inside the
dynamo region; see \Fig{bphase}.
Future studies will include the near-surface shear layer,
which has been suspected to play an important role in producing
equatorward migration \citep{B05}. 
This would also help reproducing the observed phase relation. 

\subsection{Relation between kinetic and magnetic energies}

\begin{figure}[!t]
\begin{center}
\includegraphics[width=\columnwidth]{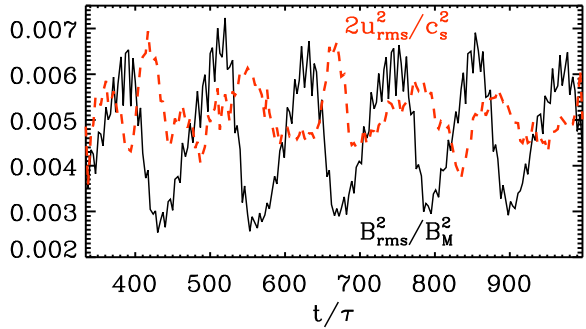}
\end{center}\caption[]{
Phase relations between the magnetic field and the velocity in the
dynamo region.
The magnetic field is plotted as $\Brms^2$, normalized with the
equipartition field of the sound speed, $B^2_M$ (=$\mu_0\rho c^2_s$)
as a solid and black line.
The rms velocity, normalized by the sound speed $\cs$, is plotted as a
dashed red line, and has been smoothed over 5 neighboring data points
to make it more legible.
Taken from Run~A.
}
\label{bu}
\end{figure}

Next we investigate the relation between the rms values of the magnetic
field and the velocity.
Both quantities are oscillating in time with a typical period of 200 turnover
times.
In \Fig{bu} we compare the time evolution of the magnetic field strength
and the rms velocity.
The magnetic field is calculated in the dynamo
region and normalized to the thermal equipartition field strength.
The phase difference between the two is slightly less than $\pi$ within the
dynamo region. 
This basically shows that the magnetic field quenches the turbulence.

\subsection{Density variations}

\begin{figure}[t!]
\begin{center}
\includegraphics[width=\columnwidth]{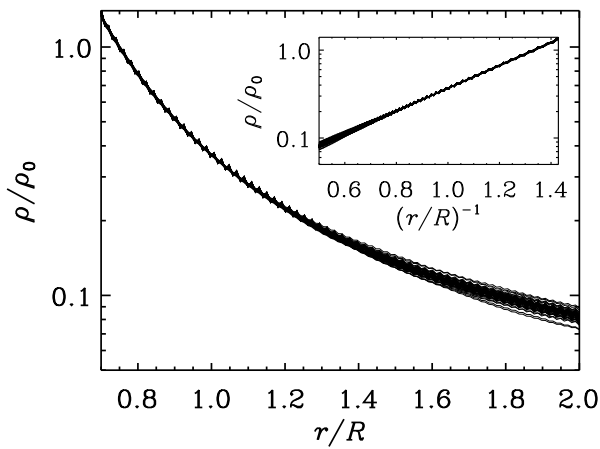}
\end{center}\caption[]{
Radial dependence of density overplotted at different times.
In the inset is the linear behavior of the logarithmic density
$\log{\rho/\rho_0}$ to the inverse of the radius shown.
Taken from Run~A.
}
\label{rho}
\end{figure}

The density is stratified in radius and varies by over an
order of magnitude.
For all the runs listed in \Tab{Summary}
the density fluctuates about the hydrostatic equilibrium
value, $\rho\approx\rho_0\exp(GM/r\cs^2)$.
The relative fluctuations are of comparable strength
at all radii; see \Fig{rho}.

\subsection{Magnetic field outside the turbulence zone}

\begin{figure}[t!]
\begin{center}
\includegraphics[width=\columnwidth]{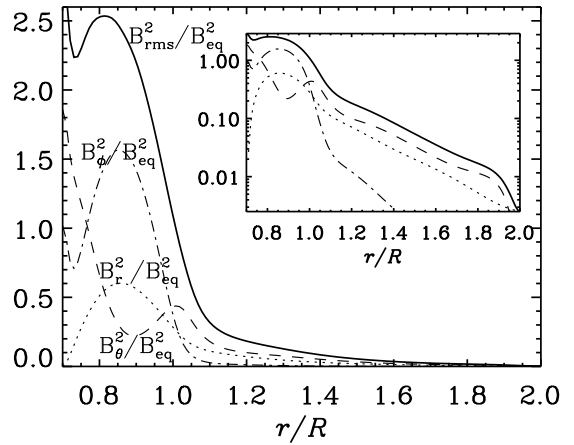}
\end{center}\caption[]{
Radial dependence of the mean squared magnetic field, $\Brms^2$ (solid line),
compared with those of $B_r$ (dotted), $B_\theta$ (dashed),
and $B_\phi$ (dash-dotted).
All quantities are averaged over 13 dynamo cycles.
The inset shows the same quantities in a logarithmic representation.
Taken from Run~A.
}
\label{brmsr}
\end{figure}

The magnetic field averaged over the entire domain is more than 5 times
smaller than in the turbulence zone.
In \Fig{brmsr} we show that the magnetic field is
concentrated to the turbulence zone and $\BB^2$ drops approximately
exponentially
with a scale height of about $0.23R$ in the outer parts for $r>R$.
The toroidal component of the magnetic field is dominant in the turbulence layer,
but does not play a significant role in the outer part.
By contrast, the radial field is weak in the inner
parts and dominates in the outer.

\begin{figure*}[t!]
\begin{center}
\includegraphics[width=3.3cm]{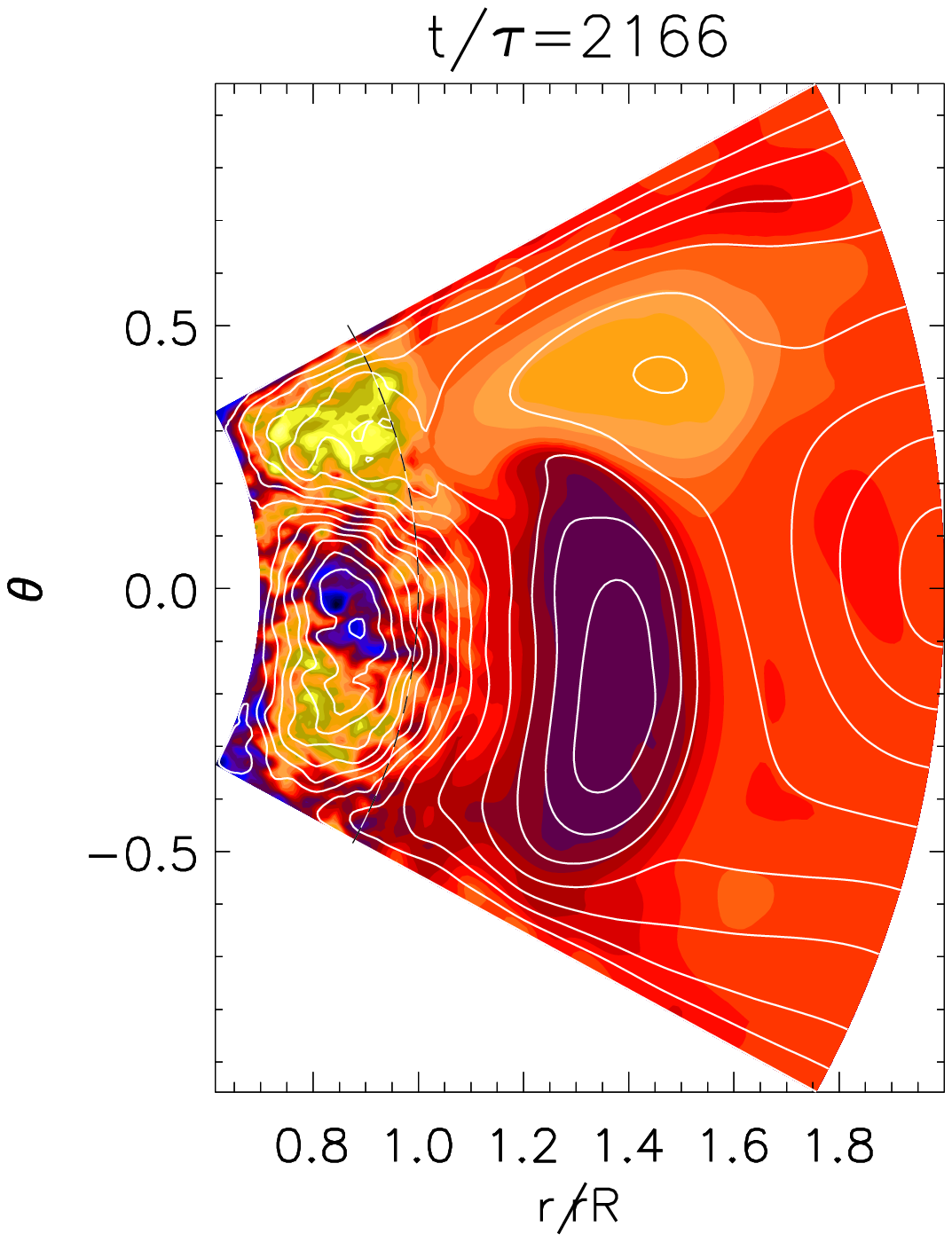}
\includegraphics[width=3.3cm]{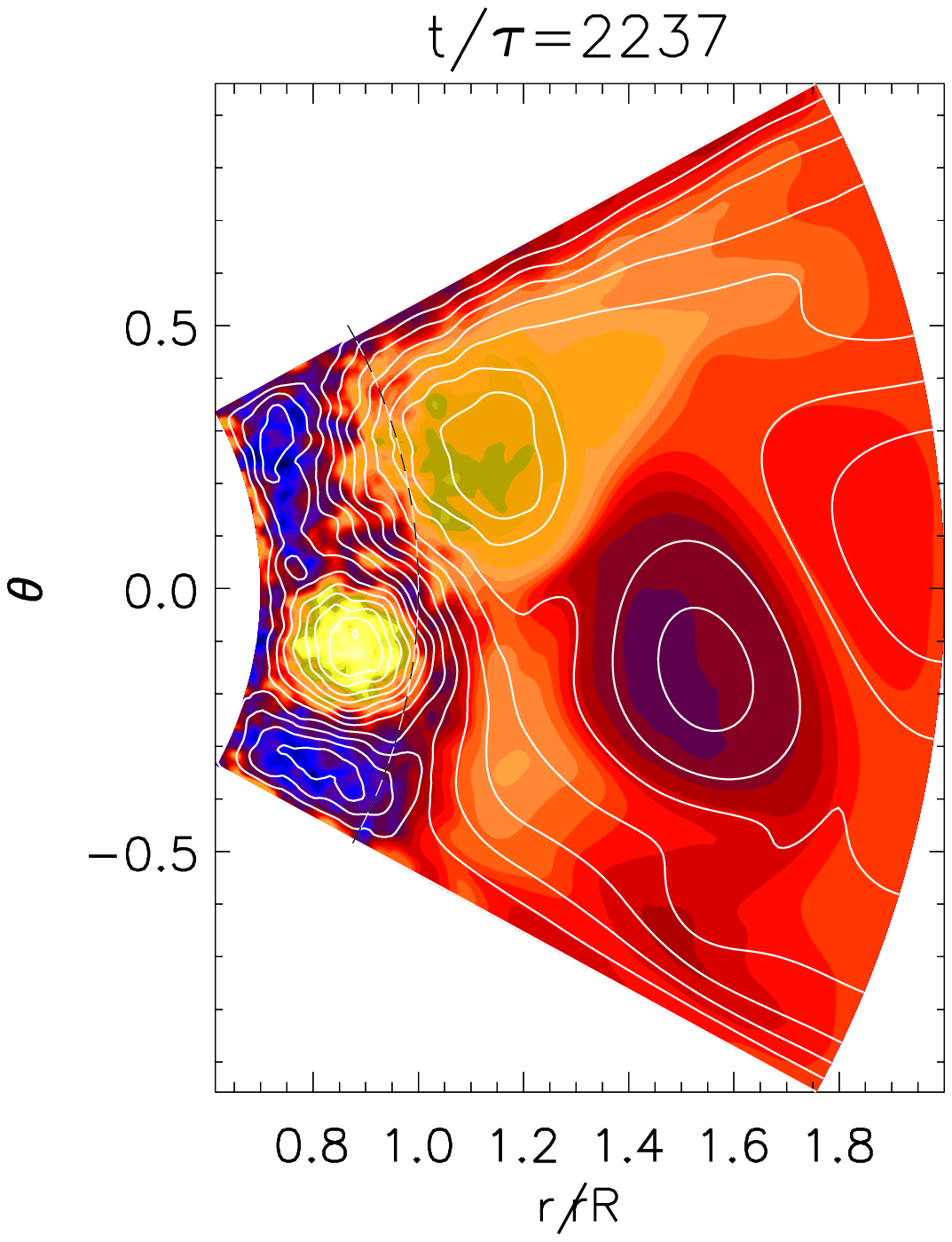}
\includegraphics[width=3.3cm]{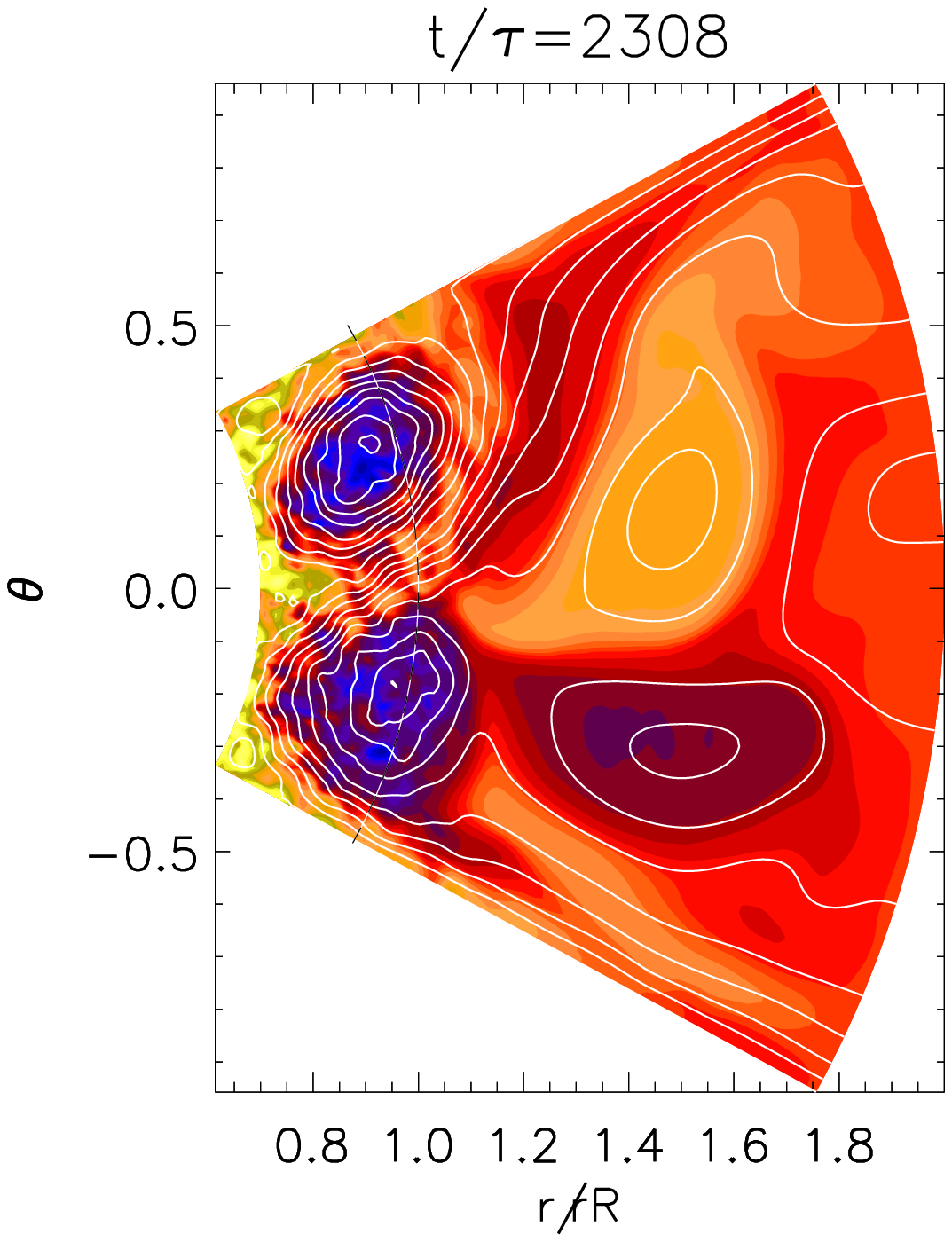}
\includegraphics[width=3.3cm]{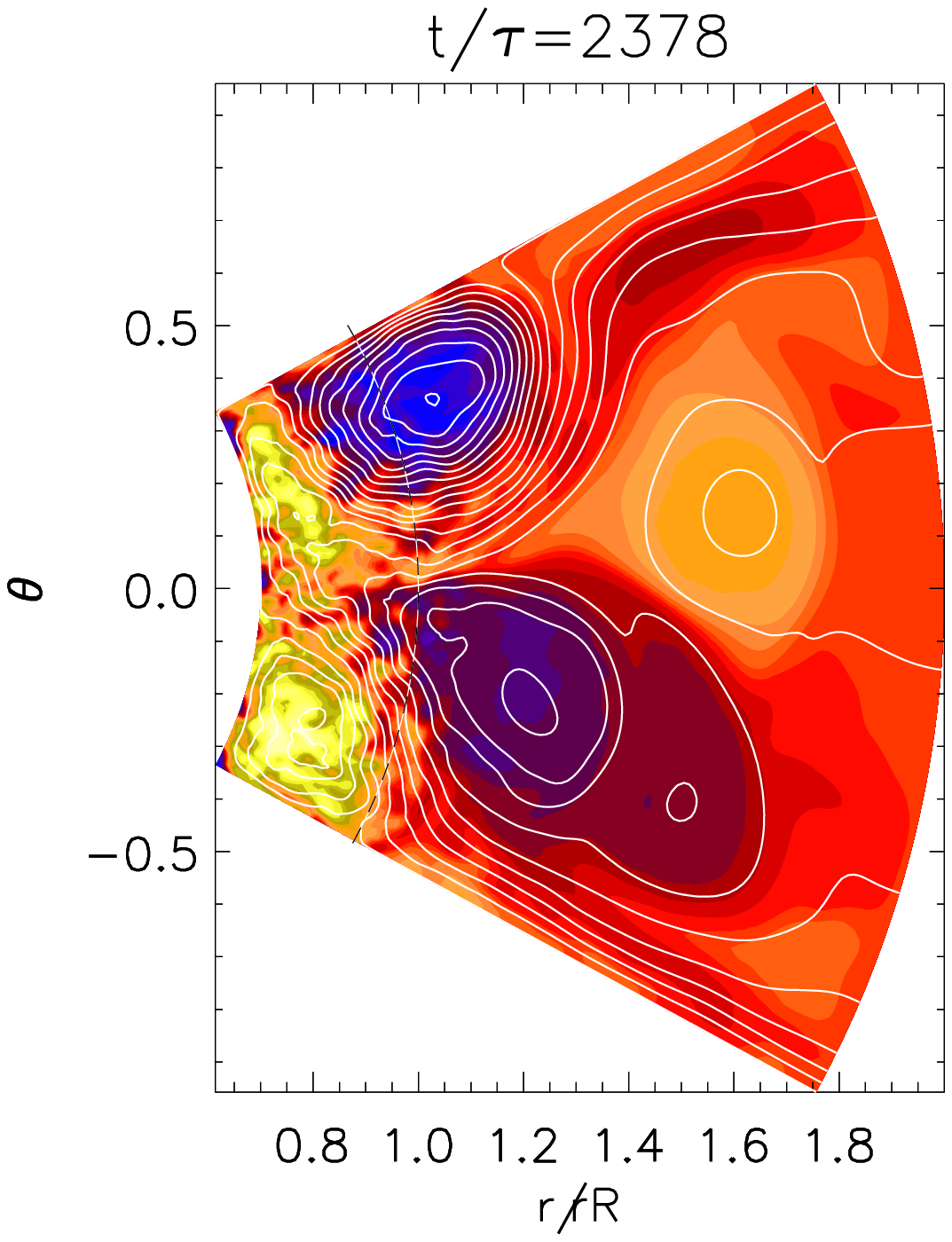}
\includegraphics[width=3.3cm]{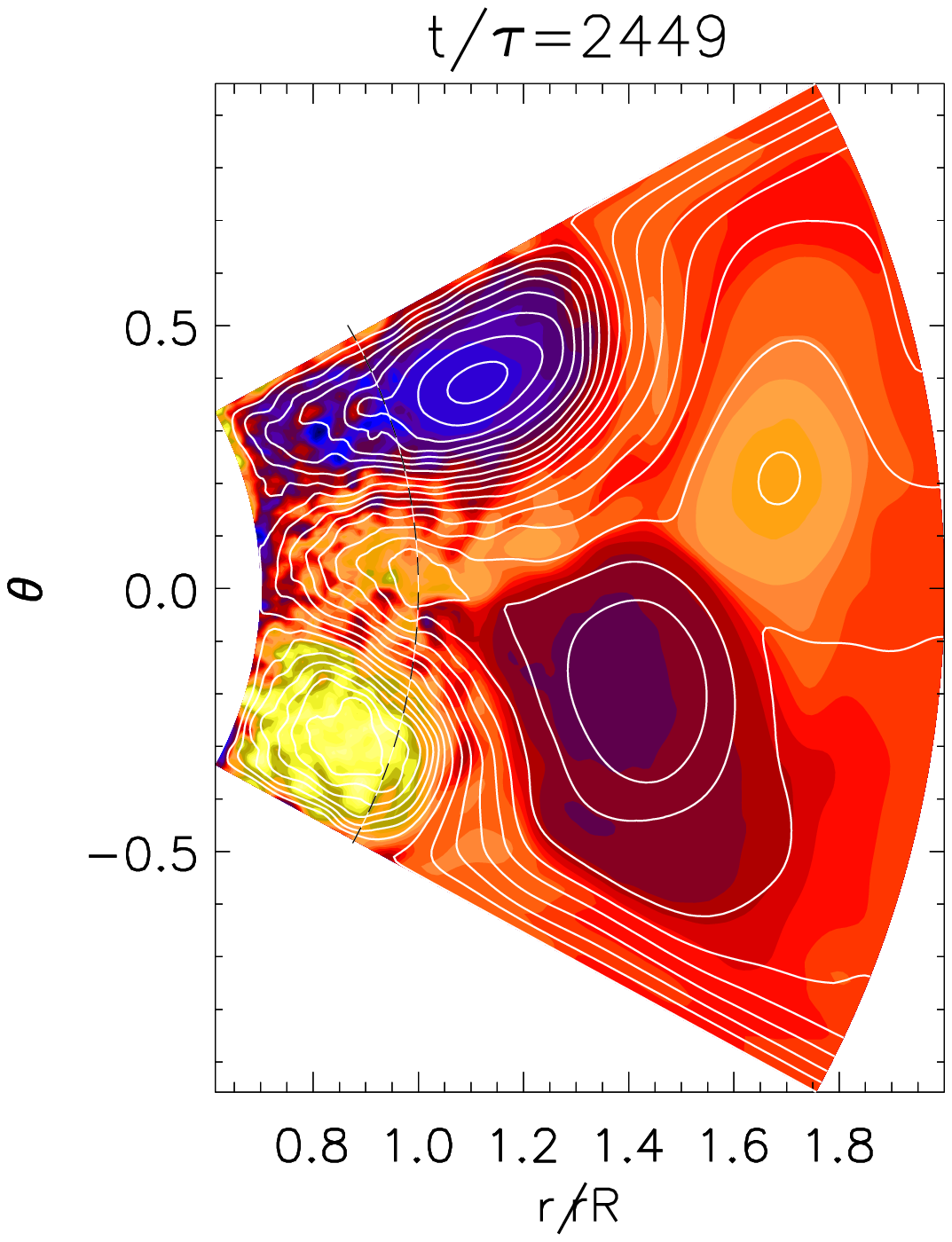}
\end{center}\caption[]{
Time series of formation of plasmoid ejections in spherical
coordinates.
Contours of $r\sin{\theta}\meanA_\phi$ are shown together
with a color-scale representation of $\meanB_\phi$; dark
blue stands for negative and light yellow for positive values.
The contours of $r\sin{\theta}\meanA_\phi$ correspond to
field lines of $\overline{\BB}$ in the $r\theta$ plane. 
The dashed horizontal lines show the location of the surface at
$r=R$.
Taken from Run D.
}
\label{A_sph}
\end{figure*}

Magnetic structures emerge through the surface
and create field line concentrations
that reconnect, separate, and rise to the outer boundary of the simulation
domain.
This dynamical evolution is seen in a sequence of field line
images in \Fig{A_sph}, where field lines of the mean field are
shown as contours of $r\sin{\theta}\meanA_\phi$ and
colors represent $\meanB_\phi$.
Prior to a plasmoid ejection we see a convergence of antiparallel radial
field lines, which then reconnect such that the newly reconnected field
lines move away from the reconnection site.
The actual reconnection seems to happen much faster than the subsequent
ejection.

In the outer layers, the magnetic field emerges as large
structures that correlate with reconnection events of magnetic fields.
In \cite{Rust94} such phenomena have been described as {\it magnetic clouds}.
We find recurrent ejections of magnetic field lines with concentrations and
reconnection events, but the occurrence of structures such as magnetic
clouds does not happen completely regularly, i.e., these structured
events would be difficult to predict.

\begin{figure}[t!]
\begin{center}
\includegraphics[width=0.8\columnwidth]{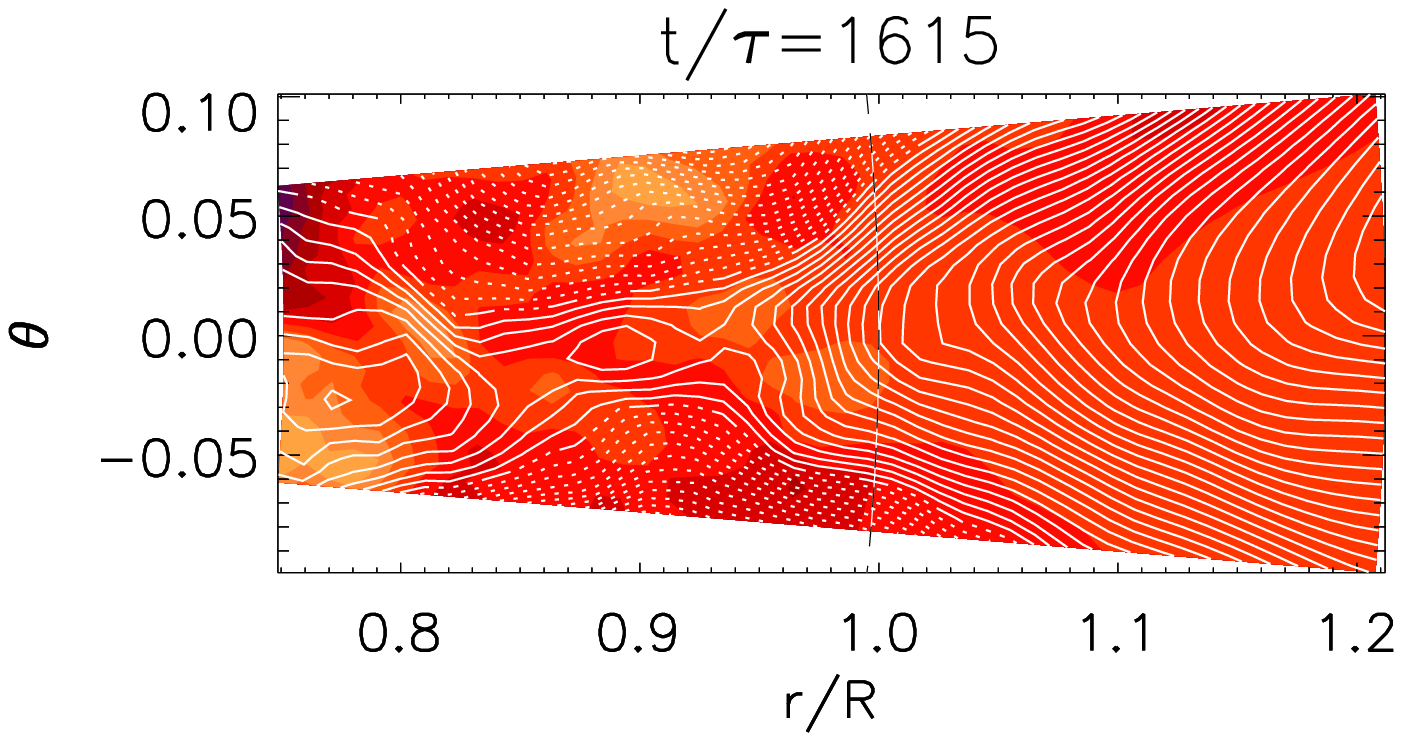}
\includegraphics[width=0.8\columnwidth]{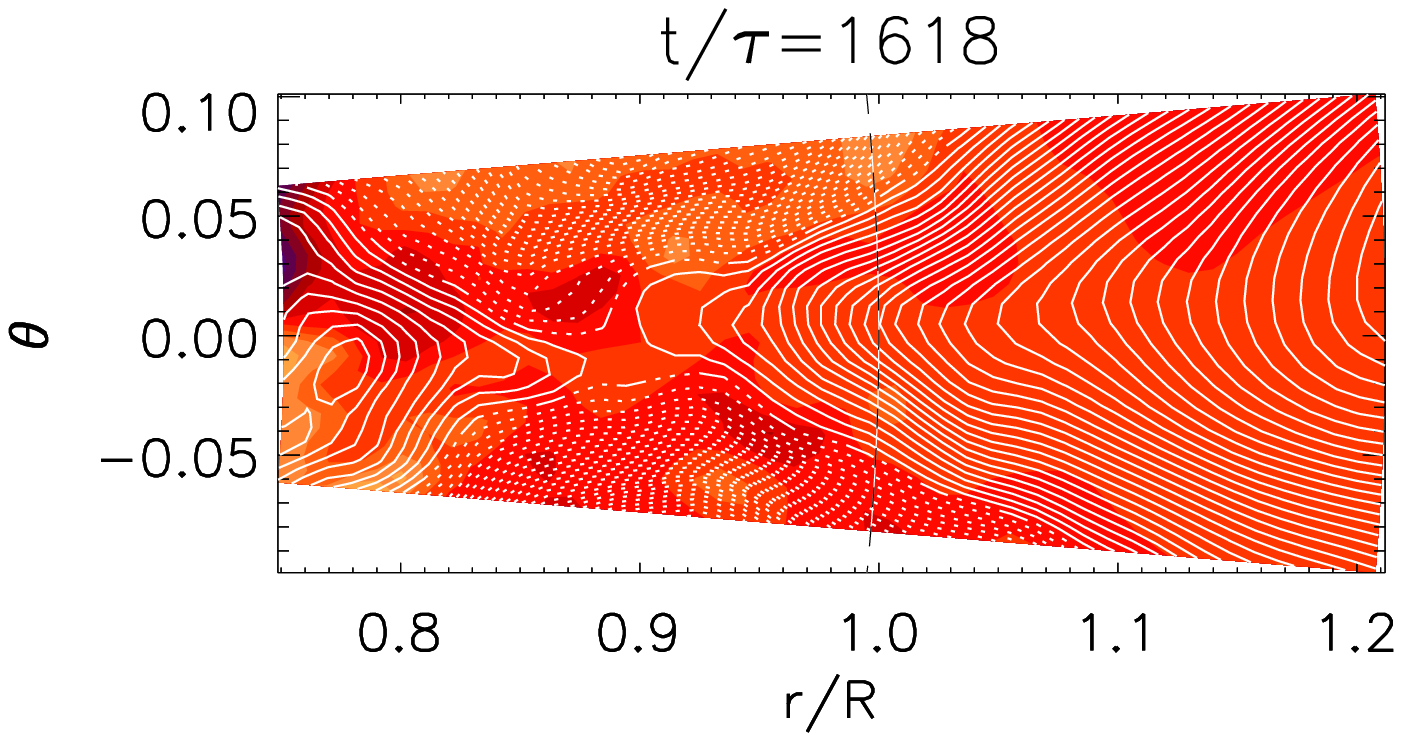}
\includegraphics[width=0.8\columnwidth]{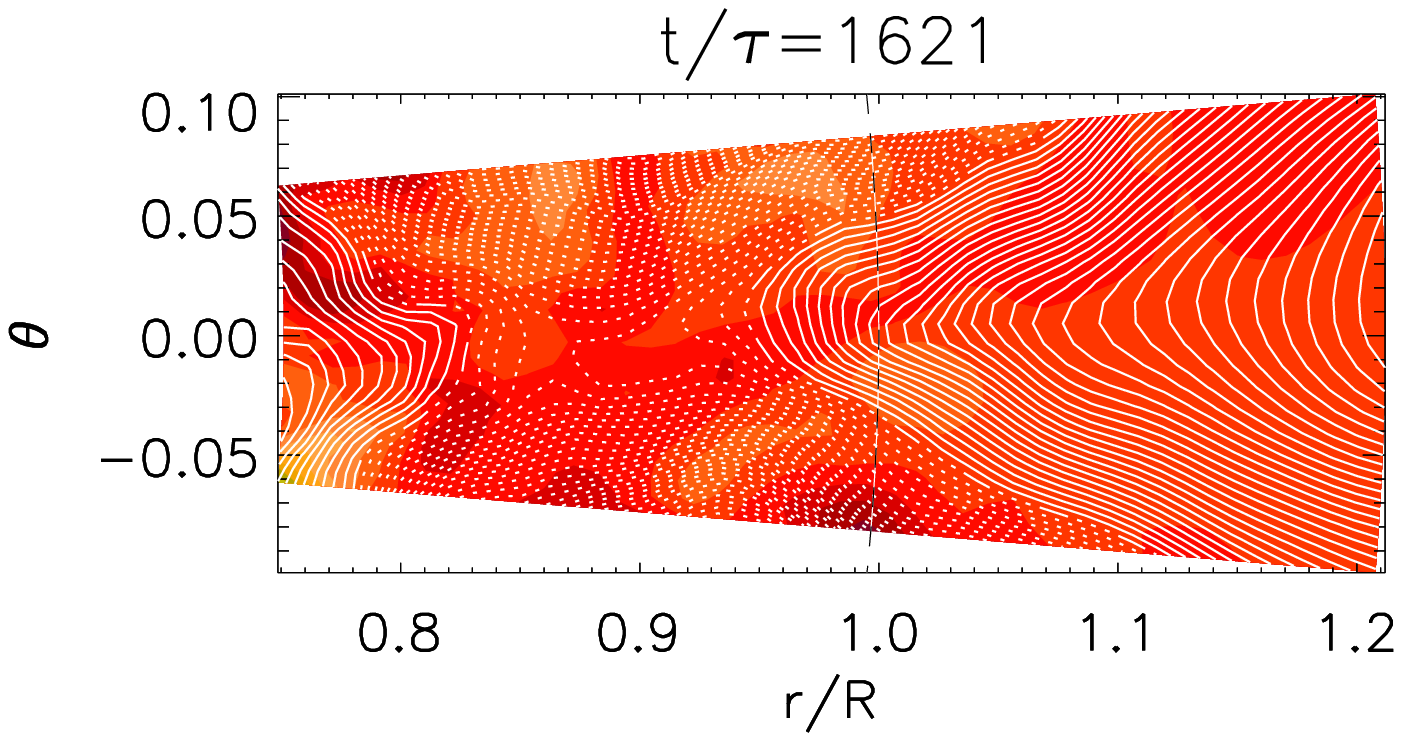}
\end{center}\caption[]{
Time series of a reconnection event in an {\sf X}-point as a close-up view.
Contours of $r\sin{\theta}\meanA_\phi$ are shown together
with a color-scale representation of $\meanB_\phi$; dark
blue stands for negative and light yellow for positive values.
The contours of $r\sin{\theta}\meanA_\phi$ correspond to
field lines of $\overline{\BB}$ in the $r\theta$ plane, where
solid lines represent counter-clockwise magnetic field lines and the dash ones
clockwise.
The dashed vertical lines show the location of the surface at $r=R$.
Taken from Run D.
}
\label{recon}
\end{figure}

In \Fig{recon} we show a close-up of the magnetic field.
A configuration resembling a reconnection event is clearly seen.
Here, the contours represent magnetic field lines with solid and dashed
lines denoting counter-clockwise and clockwise orientations,
respectively.
The solid antiparallel field lines reconnect around $r=0.9R$ and
separate to the left and to the right.
On the right-hand side, a plasmoid has formed,
which is eventually ejected. 
This plasmoid appears as a CME-like ejection in the first panel of
\Fig{jb_sph}.
These plasmoids, as seen more clearly in \Figs{A_sph}{reconII},
appear as a concentration of field lines that propagate outwards.
The fact that reconnection happens predominantly in the upper parts of
the turbulence zone suggests that turbulence is needed to enable fast
reconnection \citep{LV99}.

\begin{figure}[t!]
\begin{center}
\includegraphics[width=6cm]{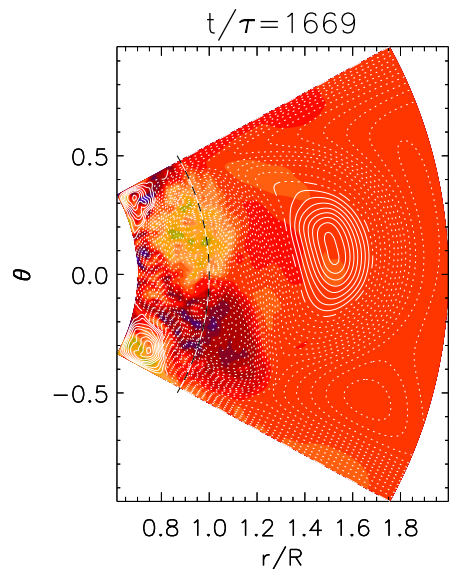}
\end{center}\caption[]{
Magnetic field configuration at the time of a ejection.
Contours of $r\sin{\theta}\meanA_\phi$ are shown together
with a color-scale representation of $\meanB_\phi$; dark
blue stands for negative and light yellow for positive values.
The contours of $r\sin{\theta}\meanA_\phi$ correspond to
field lines of $\overline{\BB}$ in the $r\theta$ plane, where the
solid lines represent counter-clockwise magnetic field lines and the dash ones
clockwise.
The dashed vertical lines show the location of the surface at $r=R$.
Taken from Run~D.
}
\label{reconII}
\end{figure}

Additionally, we find reconnection as a result of the interaction
between ejections.
As plotted in \Fig{reconII}, the field lines of two subsequent events
have the opposite field line direction, which can then interact in the
outer layers.
Comparison with the first panel of \Fig{jb_sph} shows that the current
helicity has a correlation with the separatrices of the two polarities
of the field lines.
We also find that in the interaction region the field lines
have high density and are more strongly concentrated.

\subsection{Current helicity}

\begin{figure*}[t!]\begin{center}
\includegraphics[width=3.3cm]{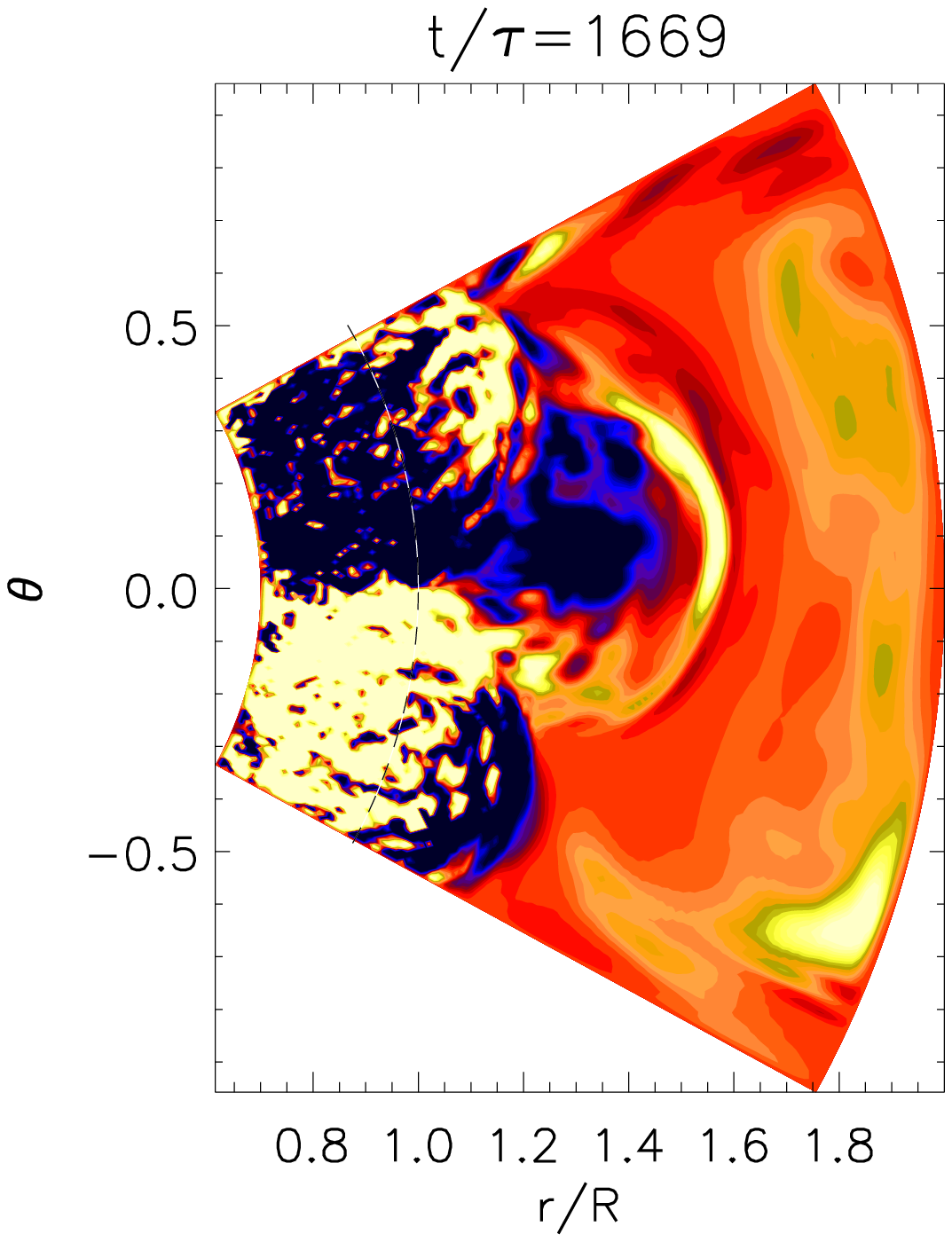}
\includegraphics[width=3.3cm]{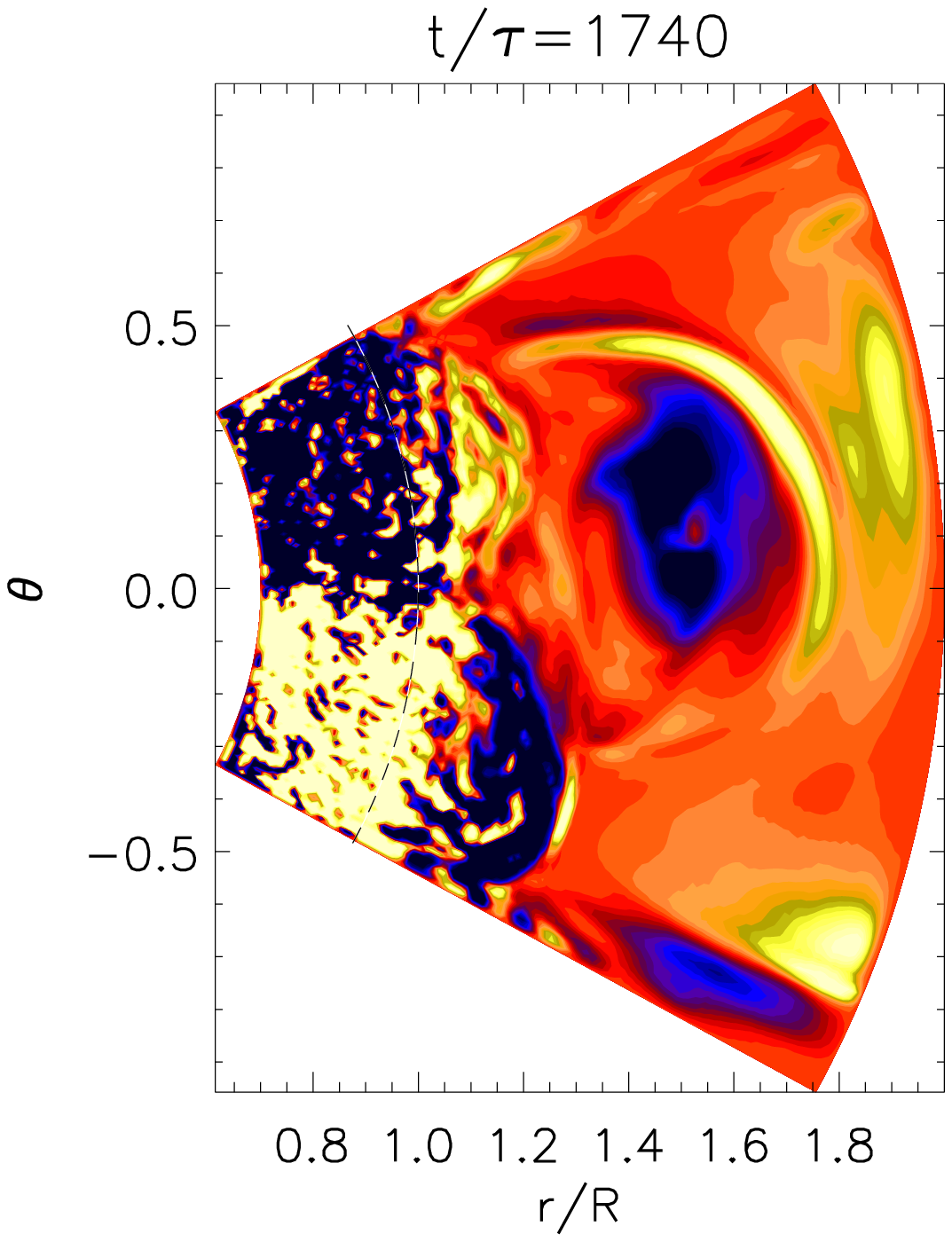}
\includegraphics[width=3.3cm]{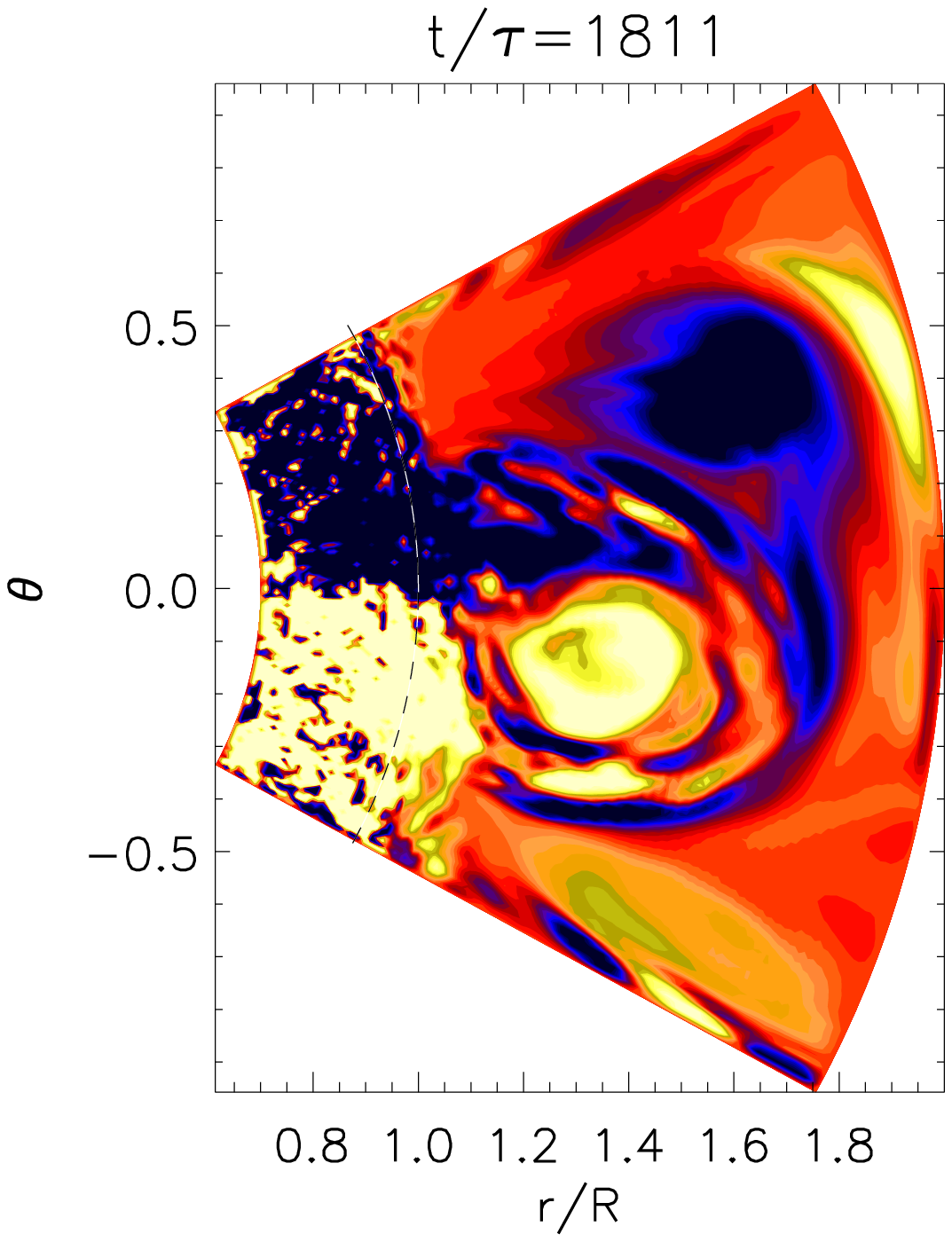}
\includegraphics[width=3.3cm]{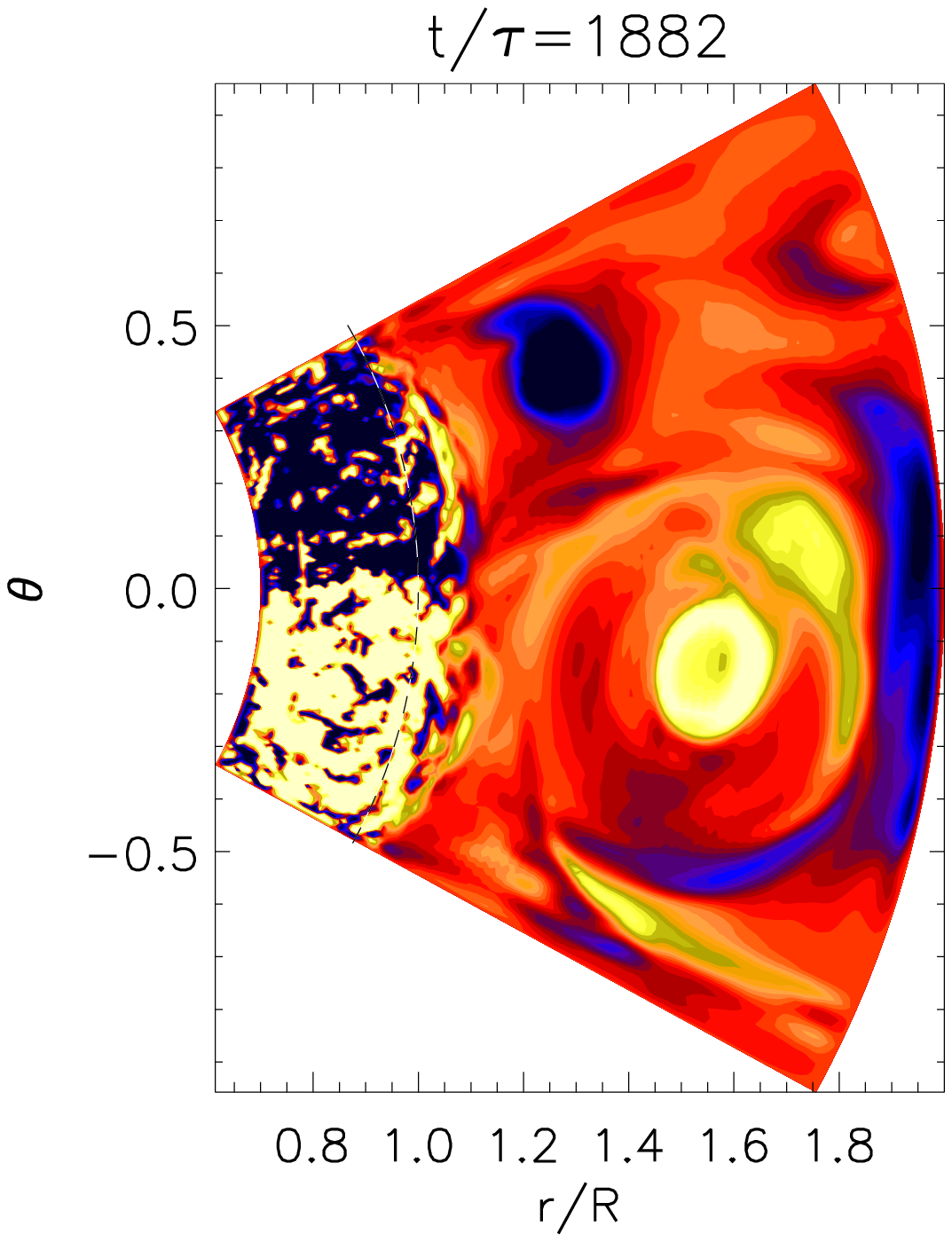}
\includegraphics[width=3.3cm]{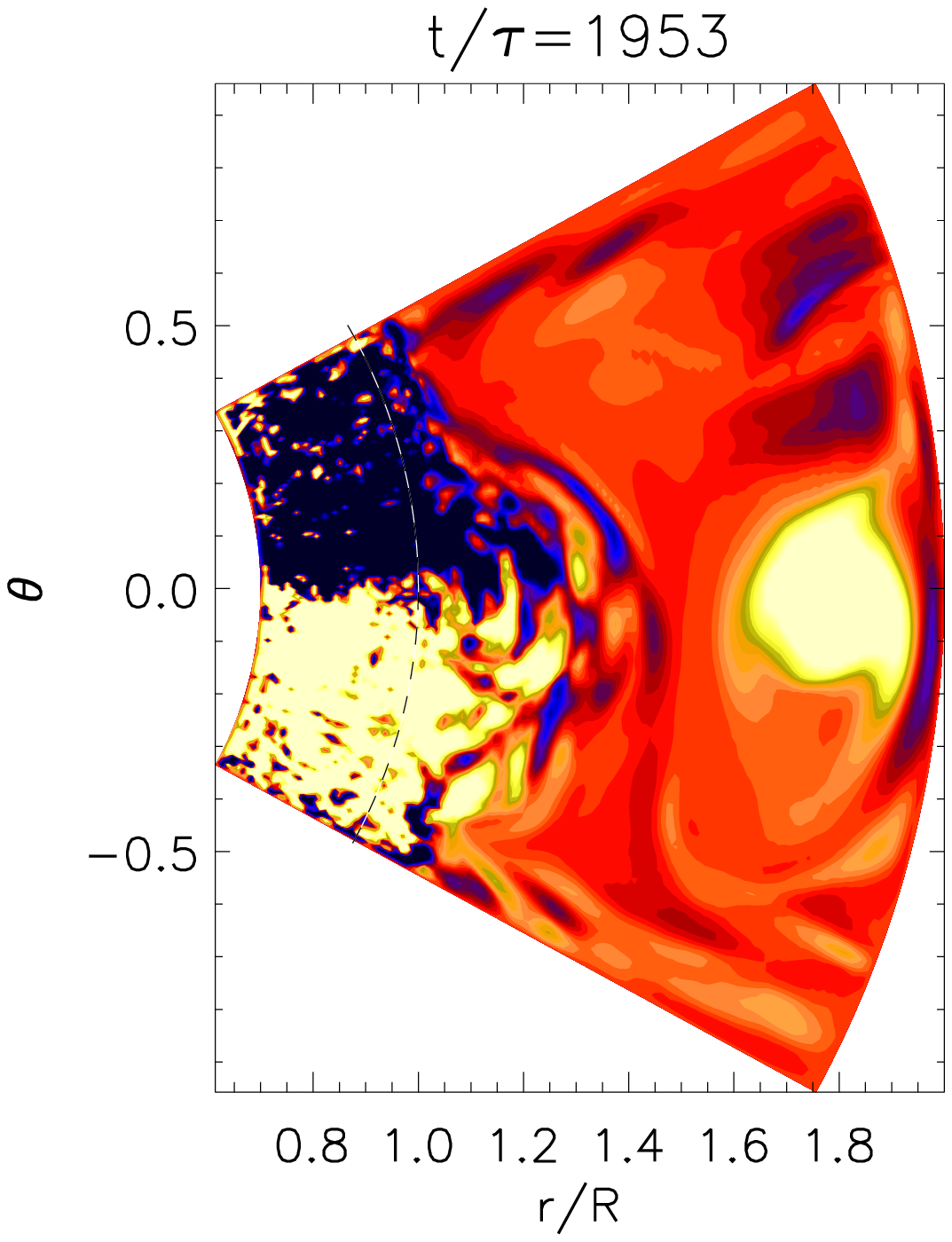}
\end{center}\caption[]{
Time series of coronal ejections in spherical coordinates.
The normalized current helicity,
$\mu_0 R\,\overline{\JJ\cdot\BB}/\bra{\overline{\BB^2}}_t$, is shown in a
color-scale representation for different times; dark blue stands for
negative and light yellow for positive values.
The dashed horizontal lines show the location of the surface at $r=R$.
Taken from Run~D.
The temporal evolution is shown in a movie available as online
material.
}
\label{jb_sph}
\end{figure*}

\begin{figure}[t!]
\begin{center}
\includegraphics[width=\columnwidth]{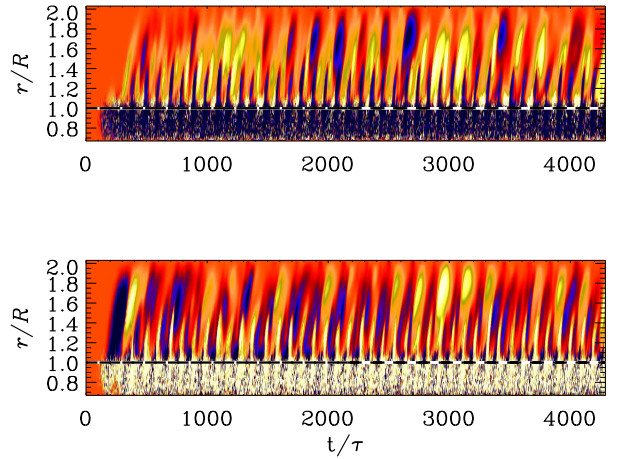}
\end{center}\caption[]{
Dependence of the dimensionless ratio
$\mu_0 R\,\overline{\JJ\cdot\BB} / \bra{\overline{\BB^2}}_t$
on time $t/\tau$ and radius $r$ in terms of the solar radius. 
The top panel
shows a narrow band in $\theta$ in the northern hemisphere and
the bottom one
a thin band in the southern hemisphere.
In both plots we have also averaged in latitude from
$20^{\circ}$ to $28^{\circ}$.
Dark blue stands for negative and light yellow for positive values.
The dotted horizontal lines show the location of the surface at $r=R$.
}
\label{jb2}
\end{figure}

The current helicity ($\JJ\cdot\BB$) is often used as a useful proxy for the 
magnetic helicity ($\AAA\cdot\BB$) at small scales,
because, unlike magnetic helicity, it is gauge-independent.
Current helicity has also been observed in the Sun \citep{See90}
and it has been obtained from mean-field dynamo models \citep{DG01}.
In the present paper we are particularly interested in the
current helicity outside the Sun.
We normalize it by the $r$-dependent time-averaged mean squared field
to compensate for the radial decrease of $\overline{\JJ\cdot\BB}$.
In \Fig{jb2}, we have also averaged in latitude from
$20^{\circ}$ to $28^{\circ}$.
In the turbulence zone the sign of
$\overline{\JJ\cdot\BB}/\bra{\overline{\BB^2}}_t$
is the same as that of kinetic helicity which, in turn, has the same sign
as the helicity of the external forcing, i.e.\ of $\sigma$;
see \Figs{jb_sph}{jb2}.

However, to our surprise, above the surface, and separately for each
hemisphere, the signs of current helicity tend to be opposite to those
in the turbulence zone; see \Fig{jb_sph} for the panels of
$t/\tau=1669$ and $t/\tau=1740$.
To demonstrate that plasmoid ejections are recurrent phenomena,
we look at the evolution of
$\overline{\JJ\cdot\BB}/\bra{\overline{\BB^2}}_{t}$
as a function of $t$ and $r$.
This is done in \Fig{jb2} for Run~A.
It turns out that the speed of plasmoid
ejecta is about 0.45 of the rms velocity of the turbulence in the
interior region for Reynolds numbers up to 15 and about 0.3 up to 30.
To compare with the Sun, we estimate the rms velocity of the turbulence
in terms of the convective energy flux via $F\approx\rho\urms^3$.
The density of the corona is $\rho_{\rm cor}\approx10^{-12}\kg\m^{-3}$,
so our estimate would suggest
$V_{\rm ej}\approx0.3(F/\rho_{\rm cor})^{{1/3}}\approx1200\kms$,
which is somewhat above the observed speeds of $400$--$1000\kms$.
The time interval between subsequent ejections is around $100\,\tau$ for Run~A. 
As seen from \Tab{Summary}, the ejection interval is independent of
the forcing amplitude $\Af$ and increases weakly with magnetic Reynolds
number, but it seems to be still comparable to half the dynamo cycle
period, i.e., $\Delta\tau\approx T_{\rm cyc}/2$.
This means that plasmoid ejections happen about twice each cycle. 
It is therefore not clear whether such a result can be extrapolated
to the real Sun.

In our simulations, we find the ejections to have the shape of the
characteristic three-part structure that is observed in real CMEs.
This is particularly clear in \Fig{jb_sph}, where the ejections seem to
contain three different parts.
In the center we find a ball-shaped structure consisting of one polarity
of current helicity, where at the front of the ejection a bow of
opposite polarity had formed.
In between these two structures the current helicity is close to zero
and appears as a cavity.
These three parts (prominence, cavity, and front) are described by
\cite{Low:1996} for CMEs and are generally referred to as
`three-part structure'.
The basic shape of the ejection is independent of the used forcing
amplitudes and the kinetic and magnetic Reynolds numbers.
It should, however, be noted that the three-part structure of the ejections
becomes clearer at magnetic higher Reynolds numbers
(e.g., for Runs~D and G compared with Run~A, for example).
In the Sun, the plasma is confined to loops of magnetic field with
flows
along field lines due to the low plasma beta in the solar corona.
This is also seen in our simulations displayed in \Fig{jb_sph}, where
the ejections follow field lines and appear to create loop-like
structures.
An animation of the detailed time evolution of the CME-like structures
emerging recurrently into the solar corona is available in the on-line
edition (see \Fig{jb_sph})\footnote{The movie is also available at
  \url{http://www.youtube.com/watch?v=aR-PgxQyP24}.}.
However, since our choice of boundary conditions does not allow mass flux
at the outer boundary, no plasma can actually leave the domain.
The recurrent nature of the plasmoid ejections found here and in
\cite{WB10} is not yet well understood.
In contrast to \cite{WB10}, where there are no strong oscillations
present, here the ejections seem to correlate with the
dynamo cycle.
In each hemisphere of the turbulence zone a magnetic field is created
with different polarity.
After they have migrated to the equator, they merge and produce an
ejection.
However, comparing with the results of \cite{WB10}, which are similar to
those in the present paper, the oscillation cannot be the only explanation
for the recurrence of the ejections.
As we have seen in \Fig{jb2}, these events export magnetic helicity out
of the domain.
For the dynamo, on the other hand, magnetic helicity losses play a
role only in the nonlinear stage.
It is therefore conceivable that the regular occurrence of plasmoid
ejections is connected with nonlinear relaxation oscillations rather
than with the dynamo cycle which is essentially a linear phenomenon.
This is also suggested by the results of \cite{WB10}, where recurrent
ejections occur without oscillations of the large-scale field.

\begin{figure}[t!]
\begin{center}
\includegraphics[width=\columnwidth]{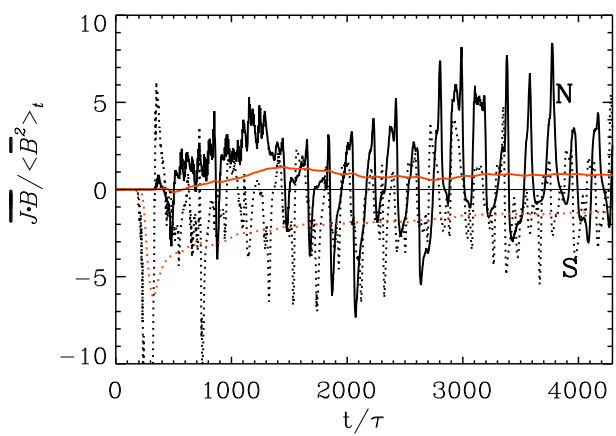}
\end{center}\caption[]{
Dependence of the dimensionless ratio
$\mu_0 R\,\overline{\JJ\cdot\BB} / \bra{\overline{\BB^2}}_t$
on time $t/\tau$ at radius $r=1.5\,R$ and $28^{\circ}$ latitude.
The solid line stands for the northern hemisphere and the dotted for
the southern hemisphere.
The red lines represent the cumulative mean for each hemisphere.
}
\label{jbP}
\end{figure}

From \Figs{jb_sph}{jb2} we conclude that 
in each hemisphere the sign of current helicity
outside the turbulence zone is mostly opposite to that inside
the turbulence zone. 
A stronger trend is shown in the cumulative mean of current helicity
over time.
This is shown in \Fig{jbP}, where we plot
the time evolution of the $\phi$ averaged current helicity
at $r=1.5\,R$ and $28^{\circ}$ latitude, which is a safe
distance away from the outer $r$ and $\theta$ boundaries so as not to
perturb our results, which should thus give a reasonable representation of
the outer layers.
For the northern hemisphere the current helicity (solid black line) and the
cumulative mean (solid red line) show positive values and for the
southern hemisphere (dotted lines) negative values.
This agrees with results of \cite{BCC09}, where the magnetic helicity
of the field in the exterior has the opposite sign than in the interior.

\begin{figure}[t!]
\begin{center}
\includegraphics[width=5cm]{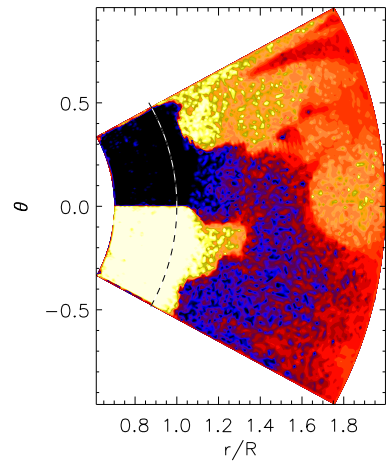}
\end{center}\caption[]{
Current helicity averaged over 3900 turnover times.
Legend is the same as in \Fig{jb_sph}.
Dark blue corresponds to negative values, while the light yellow corresponds
to positive value.
Taken from Run~D.
}
\label{jbt}
\end{figure}

To investigate whether the sign of the current helicity
is different in the turbulence zone and in the outer parts, we show
in \Fig{jbt} the azimuthally and time-averaged current helicity as a
function of radius and colatitude.
It turns out that, even though we have averaged the result over several
thousand turnover times, the hemispheric sign rule of current helicity
is still only approximately obeyed in the outer layers---even though it is
nearly perfectly obeyed in the turbulence zone.
Nevertheless, there remains substantial uncertainty, especially near
the equator.
This suggests that meaningful statements about magnetic and current
helicities in the solar wind can only be made after averaging over
sufficiently long stretches of time.

\subsection{Magnetic helicity fluxes}

\begin{figure}[t!]
\begin{center}
\includegraphics[width=\columnwidth]{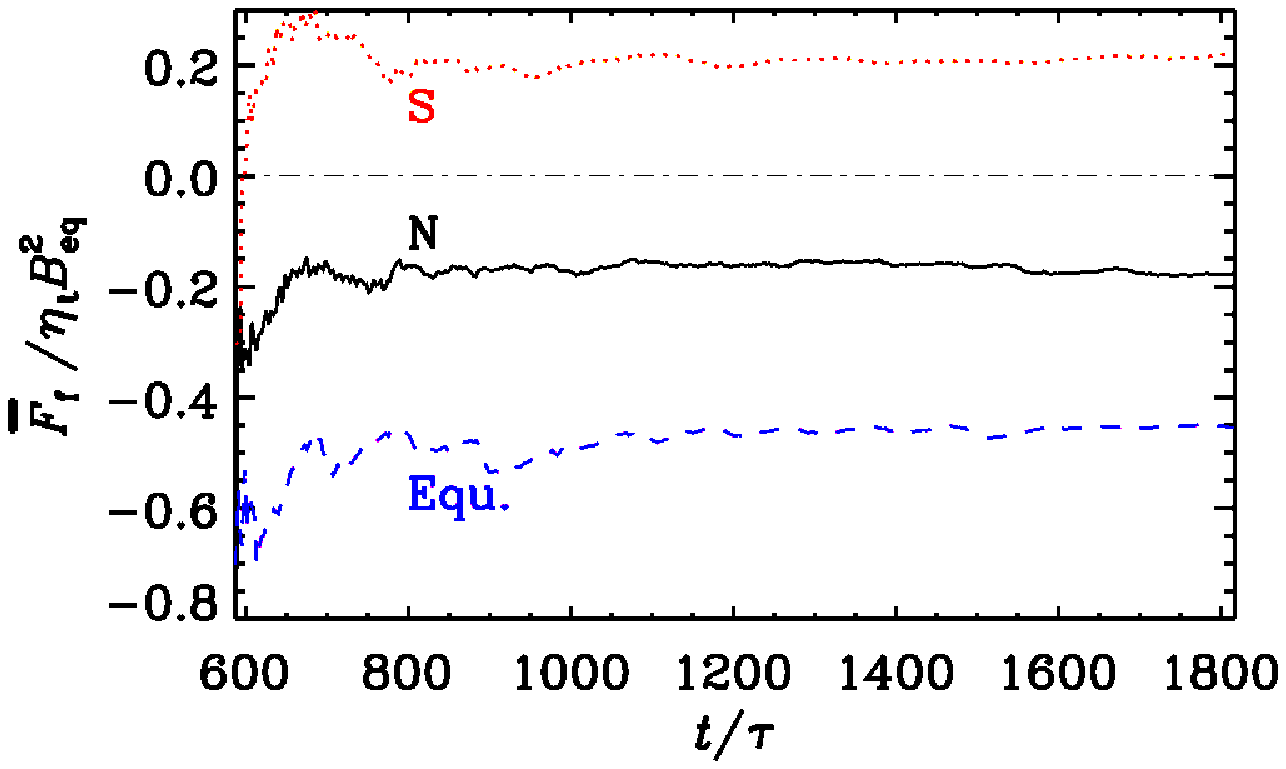}
\end{center}\caption[]{
Cumulative mean of the
time evolution of the magnetic helicity flux of the small-scale
field, $\meanFFf=\overline{\ee \times \aaaa}$,
normalized by $\etat \Beq^2$, where $\etat\approx\etatz\equiv\urms/3\kf$
was defined in \Sec{Dynamo_TZ}.
Here, the mean of magnetic helicity flux out through
the surface of the northern hemisphere (black) is shown,
together with that through the southern hemisphere (dotted red), and the
equator (dashed blue).
}
\label{helfl}
\end{figure}

In view of astrophysical dynamo theory it is important to understand the
amount of magnetic helicity that can be exported from the system.
Of particular interest here is the magnetic helicity associated with
the small-scale magnetic field.
Under the assumption of scale separation, this quantity is
gauge-independent \citep{SB06}, so we can express it in any gauge.
This has been verified in simulations both with an equator \citep{Mitra10b}
and without \citep{HB10}.
Here, we compute the magnetic helicity flux associated with the small-scale
field by subtracting that of the azimuthally averaged field from that
of the total field, i.e.,
\EQ
\overline{\ee\times\aaaa}=
\overline{\EE\times\AAA}-\meanEE\times\meanAA,
\EN
where $\EE=\mu_0\eta\JJ-\UU\times\BB$ is the electric field.
This is also the way how the magnetic helicity flux from the small-scale
field was computed in \cite{HB10}, where the magnetic helicity flux from
the total and large-scale fields was found to be gauge-dependent, but
that from the small-scale field was not.
In \Fig{helfl} we compare the flux of magnetic helicity
of the small-scale field across the outer
surfaces in the northern and southern hemispheres with that through
the equator.
It turns out that a major part of the flux goes through the equator.
The part of the magnetic helicity flux that goes through the surface is
about 20\% of $\etat\Beq^2$.
However, the magnetic helicity flux should primarily depend on
$\meanBB$ rather than $\Beq$.
In the present simulations, where the dynamo works with a fully
helical field, the two are comparable.
This is not the case in the Sun, where the estimated field strength
is expected to be about $300\G$ \citep{B09}.
Thus, to compare with the Sun, a more reasonable guess for the magnetic
helicity flux would be about 20\% of $\etat\meanBB^2$.
Integrating this over one hemisphere and multiplying this with the 11
year cycle time, we find the total magnetic helicity loss to be
$2\pi R^2\etat\meanBB^2 T_{\rm cyc}$, which corresponds to
$5\times10^{47}\Mx^2$ if we use $\etat=3\times10^{12}\cm^2\s^{-1}$.
This value exceeds the estimated upper limit for the solar dynamo
of about $10^{46}\Mx^2$ per cycle given by \cite{B09}.
However the estimate by \cite{B09} is based on an $\alpha\Omega$
dynamo model with $\alpha$ effect and shear that yield a period
comparable with the 11 year period of the Sun.
Therefore, the discrepancy with the present model,
where shear plays no role, should not be surprising.
Instead, it tells us that a dynamo without shear has to shed even more
magnetic helicity than one with shear.

\begin{figure}[t!]
\begin{center}
\includegraphics[width=\columnwidth]{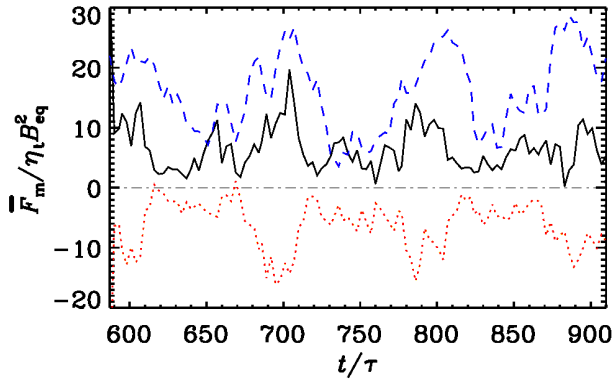}
\end{center}\caption[]{
Time evolution of the magnetic helicity flux of the large-scale
field, smooth over two data points.
Here, the mean of magnetic helicity flux out through
the surface of the northern hemisphere (black) is shown,
together with that through the southern hemisphere (dotted red), and the
equator (dashed blue).
}
\label{helflLS}
\end{figure}

The magnetic helicity flux of the large-scale field may not
{\it a priori} be gauge-invariant.
However, the system is in a statistically steady state and, in addition, 
the magnetic helicity integrated over each cycle is constant.
In that case the divergence of the magnetic helicity flux
is also gauge-invariant.
Furthermore, the shell-integrated magnetic helicity cannot
have a rotational component and is therefore uniquely defined.
In \Fig{helflLS} we plot this flux and see that its maxima
tend to occur about 50 turnover times after magnetic
field maxima; see \Figs{bphase}{bu}.
The helicity flux of the small-scale field does not show a clear behavior.
Since the ejections appear to be related to the magnetic field strength
in this way, one might conclude that the magnetic helicity flux of the
large-scale field is transported through these ejections.
This result is somewhat unexpected and deserves to be reexamined
more thoroughly in future simulations where cycle and ejection
frequencies are clearly different from each other.

 \begin{figure}[t!]
 \begin{center}
 \includegraphics[width=\columnwidth]{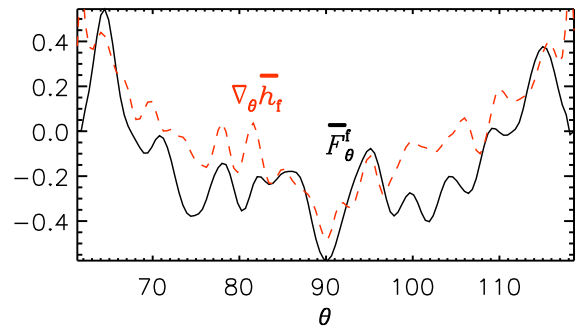}
 \end{center}\caption[]{
Dependence of the latitudinal component of the magnetic helicity flux,
$\meanF_\theta^{\rm f}$, compared with the
latitudinal gradient of the magnetic helicity density of the small-scale
field, $\nab_\theta \meanhf$, at $r/R=0.85$.
The latter agrees with the former if it is multiplied by an effective
diffusion coefficient for magnetic helicity of $\kappat\approx3\eta_{t0}$.
}
\label{kappa}
\end{figure}

Next, let us look at the magnetic helicity flux of the small-scale
field.
On earlier occasions, \cite{Mitra10b} and \cite{HB10} have been able to
describe the resulting magnetic helicity flux by a Fickian diffusion
ansatz of the form $\meanFFf=-\kappah\nab\meanhf$, where $\kappah/\etatz$
was found to be 0.3 and 0.1, respectively.
In \Fig{kappa} we show that the present data allow a similar representation,
although the uncertainty is large.
It turns out that $\kappah/\etatz$ is about 3, suggesting thus
that turbulent magnetic helicity exchange across the equator can
be rather efficient.
Such an efficient transport of magnetic helicity out of the dynamo
region is known to be beneficial for the dynamo in that it alleviates
catastrophic quenching \citep{BB03}.
In this sense, the inclusion of CME-like phenomena is not only interesting
in its own right, but it has important beneficial consequences for the
dynamo itself in that it models a more realistic outer boundary condition.

\subsection{Comparison with solar wind data}

Our results suggest a reversal of the sign of magnetic helicity
between the inner and outer parts of the computational domain.
This is in fact in agreement with recent attempts to measure
magnetic helicity in the solar wind \citep{BSBG11}.
They used the Taylor hypothesis to relate temporal fluctuations of the
magnetic field to spatial variations by using the fact that the turbulence
is swept past the space craft with the mean solar wind.
This idea can in principle also be applied to the present simulations,
provided we use the obtained mean ejection speed $V_{\rm ej}$ (see
\Tab{Summary}) for 
translating temporal variations (in $t$) into spatial ones (in $r$)
via $r=r_0-V_{\rm ej}t$.
Under the assumption of homogeneity, one can then estimate
the magnetic helicity spectrum as
$H(k)=4\,{\rm Im}(\hat{B}_\theta\hat{B}_\phi^\star)/k$;
see \cite{MGS82} and Eq.~(9) of \cite{BSBG11}.
Here, hats indicate Fourier transforms and the asterisk denotes
complex conjugation.

\begin{figure}[t!]
\begin{center}
\includegraphics[width=\columnwidth]{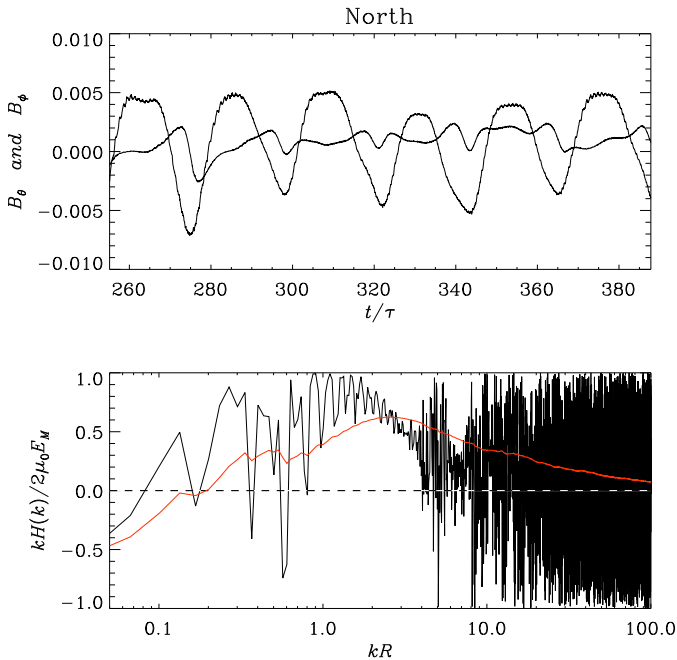}
\end{center}\caption[]{
Helicity in the northern outer atmosphere.
The values are written out at the point, $r=1.5\,R$,
$90^\circ-\theta=17^\circ$, and $\phi=9^\circ$.
{\it Top panel:} Phase relation between the toroidal $B_{\phi}$ and
poloidal $B_{\theta}$ field, plotted over time $t/\tau$.
{\it Bottom panel:} Helicity $H(k)$ is plotted over normalized wave number
$kR$.
The helicity
is calculated with the Taylor hypothesis using the
Fourier transformation of the poloidal and toroidal field.
}
\label{north}
\end{figure}

In \Figs{north}{south} we show the results
for the northern and southern hemispheres, as well as time series of
the two relevant components $B_\theta$ and $B_\phi$.
The resulting magnetic helicity spectra, normalized by $2\mu_0 E_{\rm M}/k$,
where $E_{\rm M}$ is the magnetic energy spectrum, give a quantity that is
between $-1$ and $+1$.
Note that the time traces are governed by a low frequency component
of fairly large amplitude.
In addition, there are also other components of higher frequency,
but they are harder to see.
The results suggest positive magnetic helicity in the north
and negative in the south, which would be indicative of the
helicities of the solar wind at smaller length scale.
It also agrees with the current helicities determined using
explicit evaluation in real space.
On the other hand, the Parker spiral \citep{Par58} might be responsible
for the magnetic helicity at large scales \citep{Bie87,Bie87b}.

\begin{figure}[t!]
\begin{center}
\includegraphics[width=\columnwidth]{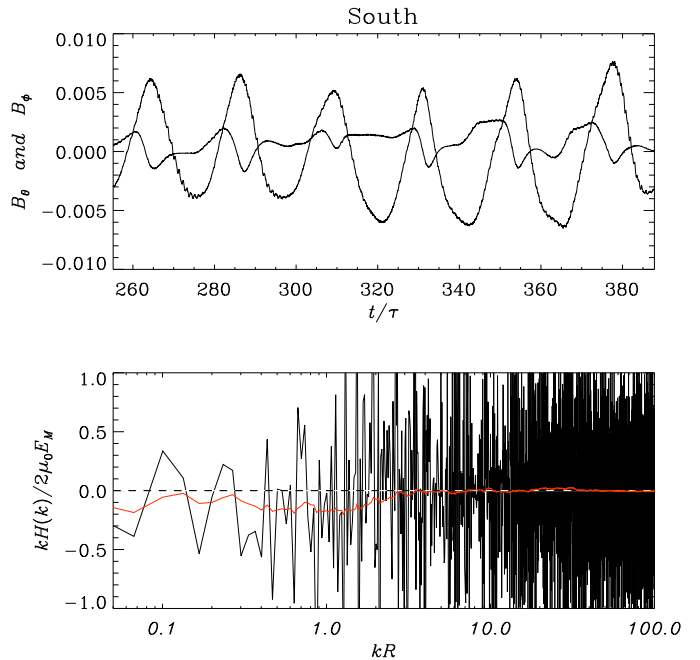}
\end{center}\caption[]{Helicity in the southern outer atmosphere.
The values are written out at the point, $r=1.5\,R$,
$90^\circ-\theta=-17^\circ$ and $\phi=8.6^\circ$.
{\it Top panel:} Phase relation between the toroidal $B_{\phi}$ and
poloidal $B_{\theta}$ field, plotted over time $t/\tau$.
{\it Bottom panel:} Helicity $H(k)$ is plotted over normalized wave number
$kR$.
The helicity
is calculated with the Taylor hypothesis using the
Fourier transformation of the poloidal and toroidal field.
}
\label{south}
\end{figure}

\section{Conclusions}

In the present work we have demonstrated that CME-like phenomena are ubiquitous
in simulations that include both a helicity-driven dynamo and a nearly
force-free exterior above it.
This was first shown in Cartesian geometry \citep{WB10} and is now also
verified for spherical geometry.
A feature common to both models is that the helical driving is confined 
to what we call the turbulence zone, which would correspond to the convection
zone in the Sun.
In contrast to the earlier work, we have now used a helical forcing for which
the kinetic helicity changes sign across the equator. 
This makes the dynamo oscillatory and displays equatorward migration of 
magnetic field \cite{Mitra10}.
More importantly, unlike our earlier work where the gas pressure was neglected
in the outer parts, it is fully retained here, because it does automatically
become small away from the surface due to the effect of gravity that is
here included too, but was neglected in the earlier Cartesian model.
The solutions shown here and those of \cite{WB10} demonstrate that this
new approach of combining a self-consistent dynamo with a corona-like
exterior is a viable one and can model successfully features that are
similar to those in the Sun.
However, our model is still not sophisticated enough for direct
quantitative comparisons. 

Of particular interest is the sign change of magnetic and current helicities
with radius.
Although similar behavior has also been seen in other Cartesian models
of \cite{BCC09}, its relevance for the Sun was unknown until evidence
for similar sign properties emerged from solar wind data \citep{BSBG11}.
In the present case we were also able to corroborate similar findings
by using the Taylor hypothesis based on the plasmoid ejection speed.
It is remarkable that this appears to be sufficient for relating
spatial and temporal fluctuations to each other.

There are many ways in which the present model can be extended and
made more realistic.
On the one hand, the assumption of isothermal stratification could
be relaxed and the increase of temperature in the corona together with the
solar wind could be modeled in a reasonably realistic way.
On the other hand, the dynamo model could be modified to include
the effects of convection and of latitudinal differential rotation.
Among other things, differential rotation would lead
to the Parker spiral \citep{Par58},
which is known to produce magnetic helicity of its own \citep{Bie87,Bie87b}.
It would then be interesting to see how this affects the magnetic helicity
distribution seen in the present model.

\begin{acknowledgements}
We thank the anonymous referee for useful suggestions and
Matthias Rheinhardt for help with the implementation.
We acknowledge the allocation of computing resources provided by the
Swedish National Allocations Committee at the Center for
Parallel Computers at the Royal Institute of Technology in
Stockholm, the National Supercomputer Centers in Link\"oping and the
High Performance Computing Center North.
This work was supported in part by
the European Research Council under the AstroDyn Research
Project No.\ 227952 and the Swedish Research Council Grant No.\ 621-2007-4064,
and the National Science Foundation under Grant No.\ NSF PHY05-51164.
\end{acknowledgements}

\end{document}